\global\def\draftcontrol{0}

   \def\versionno{Real Wall-crossing}

\catcode`\@=11

\expandafter\ifx\csname draftcontrol\endcsname\relax\global\def\draftcontrol{0} 
\fi 

{\count255=\time\divide\count255 by 60 
\xdef\hourmin{\number\count255} 
\multiply\count255 by-60\advance\count255 by\time 
\xdef\hourmin{\hourmin:\ifnum\count255<10 0\fi\the\count255}} 
\def\draftdate{\number\month/\number\day/\number\year\ \ \ \hourmin } 

\newcommand\makepapertitle{\par

  \begingroup 
    \renewcommand\thefootnote{\@fnsymbol\c@footnote}%
    \def\@makefnmark{\rlap{\@textsuperscript{\normalfont\@thefnmark}}}%
    \long\def\@makefntext##1{\parindent 1em\noindent 
            \hb@xt@1.8em{%
                \hss\@textsuperscript{\normalfont\@thefnmark}}##1}%
     \newpage 
     \global\@topnum\z@   
     \@makepapertitle 
     \thispagestyle{empty}\@thanks 
  \endgroup 
  \setcounter{footnote}{0}%
  \global\let\thanks\relax 
  \global\let\makepapertitle\relax 
  \global\let\@makepapertitle\relax 
  \global\let\@thanks\@empty 
  \global\let\@author\@empty 
  \global\let\@date\@empty 
  \global\let\@title\@empty 
  \global\let\title\relax 
  \global\let\author\relax 
  \global\let\date\relax 
  \global\let\and\relax 
  \def\version{\let\version\@version\@gobble} 
} 
\def\@makepapertitle{%
  \newpage 
   \ifnum\draftcontrol=1 {} 
   \version\versionno 
   \vskip 5.5em%
   \else 
   \hfill\hbox to 3cm {\parbox{4cm}{\@pubnum}\hss}%
   \vskip 6.5em%
   \fi 
   \begin{center}%
   \let \footnote \thanks 
      {\hskip -0\textwidth \hbox to 1\textwidth%
        {\centerline{\Large\bf{\noindent\@title}}}}%
     \vskip 2em%
     {\normalsize
       \lineskip .5em%
       \begin{tabular}[t]{c}%
         \@author 
       \end{tabular}\par}%
     \vskip 1.5em%
     {\@bstract}%
     \end{center}%
     \vfill
     \@date%
     \vskip 1.5em%
   \par 
} 

\gdef\@pubnum{} 
\def\pubnum#1{%
  \gdef\@pubnum{#1}} 

\gdef\@bstract{} 
\def\Abstract#1{%
  \gdef\@bstract{%
   \parbox{\textwidth-0pc}{%
   \centerline{\bf Abstract}\penalty1000 
   \noindent
   \renewcommand\baselinestretch{1.0} 
   {#1}}} 
} 

\gdef\@email{}
\def\email#1{%
   \gdef\@email{%
   Email: {\tt #1}}
}

\def\ps@paper{\let\@mkboth\@gobbletwo%
     \ifnum\draftcontrol=1 
        \def\@oddfoot{\hbox to \textwidth{\tiny \versionno \hfil\tiny\draftdate}%
        \hskip -\textwidth \hbox to \textwidth{\hfil\rm\thepage\hfil}}%
     \else\def\@oddfoot{\hbox to \textwidth{\hfil\rm\thepage\hfil}} 
     \fi 
     \let\@evenfoot\@oddfoot 
} 

\def\body{\clearpage 
          \pagestyle{paper} 
        } 
\newenvironment{acknowledgments}{%
\vskip 3.25ex 
\addcontentsline{toc}{section}{Acknowledgments}
\noindent {\bf Acknowledgments} 
} 

\def\@version#1{\ifnum\draftcontrol=1 
\typeout{}\typeout{#1}\typeout{} 
\vskip3mm\centerline{\hbox{\fbox{\normalsize{\tt DRAFT -- #1 -- } 
                   {\draftdate}}}}\vskip3mm 
\fi} 
\let\version\@version 
\long\def\eqlabel#1{\ifnum\draftcontrol=1 
                    \tag@false  
                    \tag*{(\theequation) \hbox to -0.2cm{\hspace{0cm}\small{#1}\hss}} 
                    \refstepcounter{equation}  
                    \edef\@currentlabel{\theequation} 
                    \ltx@label{#1}          
                    \else 
                    \label{#1} 
                    \fi 
                    } 
\let\st@bibitem\@bibitem 
\let\st@lbibitem\@lbibitem 
\ifnum\draftcontrol=1 
  \def\@bibitem#1{%
    \st@bibitem{#1}\a@@label{#1}\ignorespaces} 
  \def\@lbibitem[#1]#2{%
    \st@lbibitem[#1]{#2}\a@@label{#2}\ignorespaces} 
  \def\a@@label#1{%
    \gdef\a@lab{\smash{\normalfont\small#1}} 
    \ifvmode 
      \if@inlabel 
        \global\setbox\@labels\hbox{%
          \llap{\a@lab\let\a@lab\relax 
                \kern\@totalleftmargin\kern\marginparsep}%
          \box\@labels}%
      \fi 
    \fi} 
\fi 

\documentclass[12pt,letterpaper]{article} 

\usepackage{amsmath,bm,amsfonts,amssymb,array,calc,amsthm,rotating}
\usepackage{epsfig,psfrag} 
\usepackage{rotating}
\usepackage{amscd}
\usepackage{graphicx}
\usepackage{color}
\usepackage[colorlinks=false]{hyperref}

\renewcommand\baselinestretch{1.25} 
\renewcommand\section{\@startsection {section}{1}{\z@}%
                                   {-3.5ex \@plus -1ex \@minus -.2ex}%
                                   {2.3ex \@plus.2ex}%
                                   {\normalfont\large\bfseries}} 
\renewcommand\subsection{\@startsection{subsection}{2}{\z@}%
                                   {-3.25ex\@plus -1ex \@minus -.2ex}%
                                   {1.5ex \@plus .2ex}%
                                   {\normalfont\normalsize\bfseries}} 
\renewcommand\subsubsection{\@startsection{subsubsection}{3}{\z@}%
                                   {-3.25ex\@plus -1ex \@minus -.2ex}%
                                   {1.5ex \@plus .2ex}%
                                   {\normalfont\normalsize\it}} 
\renewcommand\paragraph{\@startsection{paragraph}{4}{\z@}%
                                   {-3.25ex\@plus -1ex \@minus -.2ex}%
                                   {1.5ex \@plus .2ex}%
                                   {\normalfont\normalsize\bf}} 
\renewcommand\subparagraph{\@startsection{subparagraph}{5}{\z@}%
                                   {-1.25ex\@plus -1ex \@minus -.2ex}%
                                   {0ex \@plus .2ex}%
                                   {\normalfont\normalsize\it}}

\numberwithin{equation}{section}

\long\def\@makecaption#1#2{%
  \vskip\abovecaptionskip
  \sbox\@tempboxa{{\bf #1:} #2}%
  \ifdim \wd\@tempboxa >\hsize
    {\small\bf #1:} {\small #2}\par
  \else
    \global \@minipagefalse
    \hb@xt@\hsize{\hfil\box\@tempboxa\hfil}%
  \fi
  \vskip\belowcaptionskip}

\renewcommand*\l@section[2]{%
  \ifnum \c@tocdepth >\z@
    \addpenalty\@secpenalty
    \addvspace{.5em \@plus\p@}%
    \setlength\@tempdima{1.5em}%
    \begingroup
      \parindent \z@ \rightskip \@pnumwidth
      \parfillskip -\@pnumwidth
      \leavevmode \bfseries
      \advance\leftskip\@tempdima
      \hskip -\leftskip
      #1\nobreak\hfil \nobreak\hb@xt@\@pnumwidth{\hss #2}\par
    \endgroup
  \fi}
\renewcommand*\l@subsection{\addvspace{.0em \@plus\p@}\@dottedtocline{2}{1.5em}{2.3em}}
\renewcommand*\l@subsubsection{\addvspace{-.2em \@plus\p@}\@dottedtocline{3}{3.8em}{3.2em}}

\def\hepth#1{\href{http://xxx.arxiv.org/abs/hep-th/#1}{{arXiv:hep-th/#1}}}

\def\math#1{\href{http://xxx.arxiv.org/abs/math/#1}{{arXiv:math/#1}}}

\def\mathag#1{\href{http://xxx.arxiv.org/abs/math.AG/#1}{{arXiv:math.ag/#1}}}

\def\arxiv#1#2{\href{http://xxx.arxiv.org/abs/#1}{{arXiv:#1 [#2]}}}

\definecolor{refcol}{rgb}{0.2,0.2,0.8}
\definecolor{eqcol}{rgb}{.6,0,0}
\definecolor{purple}{cmyk}{0,1,0,0}

\gdef\@citecolor{refcol}
\gdef\@linkcolor{eqcol}
\def\colorlinkspurple{\gdef\@urlcolor{purple}}
\def\colorlinksblue{\gdef\@urlcolor{blue}}
\def\colorlinksred{\gdef\@urlcolor{red}}

\def\ie{{\it i.e.}}

\def\cf{{\it cf.}}

\def\revise#1       {\raisebox{-0em}{\rule{3pt}{1em}}%
                     \marginpar{\raisebox{.5em}{\vrule width3pt\ 
                     \vrule width0pt height 0pt depth0.5em 
                     \hbox to 0cm{\hspace{0cm}{%
                     \parbox[t]{4em}{\raggedright\footnotesize{#1}}}\hss}}}}

\newcommand\topa[2]{\genfrac{}{}{0pt}{2}{\scriptstyle #1}{\scriptstyle #2}}

\def\sqr#1#2{{\vcenter{\vbox{\hrule height.#2pt   
 \hbox{\vrule width.#2pt height#1pt \kern#1pt 
 \vrule width.#2pt}\hrule height.#2pt}}}}

\newcommand{\beq}{\begin{equation}}
\newcommand{\eq}{\end{equation}}

\newcommand{\bracket}[3]{\left<#1\left|#2\right|#3\right>}
\newcommand{\ket}[1]{\left|#1\right>}
\newcommand{\bra}[1]{\left<#1\right|}
\newcommand{\emptypar}{\cdot}

\newcommand{\tr}{{\rm Tr}}

\newcommand{\req}[1]{(\ref{#1})}

\newcommand{\C}{\mathbb C}
\newcommand{\Z}{\mathbb Z}
\renewcommand{\P}{\mathbb P}
\renewcommand{\O}{\mathcal O}
\newcommand{\N}{\mathbb N}
\newcommand{\I}{\mathcal I}
\newcommand{\T}{\mathbb T}
\newcommand{\R}{\mathbb R}

\newcommand{\Zcal}{\mathcal Z}
\newcommand{\Pcal}{\mathcal P}
\newcommand{\Scal}{\mathcal S}
\newcommand{\Gcal}{\mathcal G}
\newcommand{\Wcal}{\mathcal W}
\newcommand{\Ncal}{\mathcal N}
\newcommand{\Mcal}{\mathcal M}

\renewcommand{\t}{\tilde}
\newcommand{\h}{\hat}
\renewcommand{\c}{\check}
\newcommand{\ii}{{\it i}}
\newcommand{\cc}{\bar}
\renewcommand{\d}{\partial}

\newcommand{\Gam}[1]{\Gamma_-\left(#1 \right)}
\newcommand{\Gamp}[1]{\Gamma'_-\left(#1 \right)}
\newcommand{\Gap}[1]{\Gamma_+\left(#1 \right)}
\newcommand{\Gapp}[1]{\Gamma'_+\left(#1 \right)}

\newcommand{\Hilb}{{\rm Hilb}}
\newcommand{\spec}{{\rm Spec}}
\newcommand{\Hom}{{\rm Hom}}

\newcommand{\inv}{\tau}

\catcode`\@=12 

\begin{document} 


\title{Wall Crossing Phenomenology of Orientifolds}

\pubnum{
arXiv:1001.5031\\
IPMU10-0016
}
\date{January 2010}

\author{
Daniel Krefl\\[0.2cm]
 IPMU, The University of Tokyo, Kashiwa, Japan
}

\Abstract{
We initiate the study of wall crossing phenomena in orientifolds of local toric Calabi-Yau 3-folds from a topological string perspective. For this purpose, we define a notion of real Donaldson-Thomas partition function at the large volume, orbifold and non-commutative point in K\"ahler moduli space. As a byproduct, we refine the constant map contribution to the partition function of the real topological string on a local toric background. We conjecture the general relation between the real large volume and real non-commutative/orbifold Donaldson-Thomas partition function of orientifolds of local toric Calabi-Yau 3-folds without compact divisors. The conjectured relation is confirmed at hand of the conifold and local $A_n$ singularity, for which we explicitly derive the real non-commutative/orbifold Donaldson-Thomas partition function combinatorially.
}

\makepapertitle

\body

\version\versionno

\vskip 1em

\tableofcontents
\newpage

\section{Introduction}
Recently, some notable progress has been achieved in understanding orientifolds in the context of topological string theory (with the orientifold acting anti-holomorphically in the A-model), both in compact \cite{Walcher:2007qp} and non-compact \cite{Krefl:2009md,Krefl:2009mw} settings (for previous studies of orientifolds in topological string theory see \cite{Sinha:2000ap,Acharya:2002ag,Bouchard:2004iu,Bouchard:2004ri}). Perhaps the most astonishing new observation is a topological version of tadpole cancellation \cite{Walcher:2007qp}. Namely, it appears that if the orientifold possesses a fixed-point locus of specific topology, only a combination of unoriented and open topological amplitudes possesses an expansion into Gopakumar-Vafa like integer invariants, \ie, can be given a BPS state counting interpretation (for followup studies of topological tadpole cancelation, see \cite{Cook:2008eu,Bonelli:2009aw}). The open topological sector needed for consistency comes from a single D-brane, which necessarily sits on the special Lagrangian 3-cycle defined by the fixed-point locus, \ie, on the O-plane. Such a D-branes is usually referred to as real brane. In the non-compact case, one should note that the D-brane defined in this fashion does not necessarily belong to the universality class of ``toric" branes constructed in \cite{Aganagic:2000gs}. The topological string on such a background with O-plane given by the real locus of an anti-holomorphic involution and a single D-brane placed on top is usually referred to as the real topological string. 

While in the B-model formulation of the real topological string on, both, compact and non-compact backgrounds still some obstacles have to be overcome in order to achieve the same degree of computability as the ordinary topological string (for the current state of art, see \cite{Krefl:2009md}), in the A-model on non-compact toric backgrounds both are now on equal footing thanks to the real vertex formalism developed in \cite{Krefl:2009md,Krefl:2009mw} as an extension of the ordinary topological vertex of \cite{Aganagic:2003db}. Besides the computational power of the real and ordinary vertex formalism (the vertex yields the full genus expansion for given degree), a remarkable aspect of the ordinary vertex is that it allows to identify topological string partition functions with the partition functions of certain combinatorial systems \cite{Okounkov:2003sp}. Thus, the vertex translates the calculation of Gromov-Witten invariants, being roughly said a count of holomorphic maps from a Riemannian surface to a target space, to a combinatorial problem. In detail, the combinatorial systems whose partition functions are equivalent to topological string partition functions are sets of 3d partitions (also known as plane partitions) with boundary conditions along the three axes given by 2d partitions (Young diagrams), where the 3d partitions are patched together along common boundaries to form a crystal like arrangement. Since a 3d partition can be seen as a monomial ideal, which in turn can be seen as an ideal sheaf on a $\C^3$ patch, an interpretation of the combinatorial system in terms of ideal sheaves is implicit. Indeed, one can show that the combinatorial systems described above compute Donaldson-Thomas invariants \cite{Iqbal:2003ds,MNOPI}. 
Hence, the vertex formalism allows to connect topological string theory with Donaldson-Thomas theory. At least for local toric backgrounds, the precise relation between the topological string (Gromov-Witten) and Donaldson-Thomas partition function is a simple change of variables \cite{MNOPI}, \ie, 
\beq
\t\Zcal(Q_i,q)= \t Z(Q_i,e^{g_s}\rightarrow -q)\,,
\eq
where we denoted the Gromov-Witten partition function without constant map contribution by $\t Z$ and the Donaldson-Thomas partition function without degree $0$ contribution by $\t \Zcal$, $Q_i:=e^{-t_i}$ refers to the set of K\"ahler parameters $t_i\in H_2(X,\Z)$, $g_s$ denotes the string coupling and $q$ is the expansion parameter of the Donaldson-Thomas partition function.  

From a physical point of view, one can see the Donaldson-Thomas partition function as a generating function counting BPS bound states of D6, D2 and D0 branes (being particles in space-time), at least for appropriate values of the B-field \cite{Denef:2007vg}. We will denote the generating function of such bound states as $Z_{BPS}$. The partition function $Z_{BPS}$ is globally defined over K\"ahler moduli space. However, BPS states exhibit wall crossing phenomena upon crossing walls of marginal stability (where BPS bound states can decay/form), hence, $Z_{BPS}$ is a discontinuous function and only at a specific region in K\"ahler moduli space there is an interpretation in terms of Donaldson-Thomas invariants. One might ask if $Z_{BPS}$ counts other mathematical invariants at other regions in moduli space. Indeed, at specific points in K\"ahler moduli space other than large volume, like for example the orbifold \cite{Young08} or non-commutative point \cite{Szendroi07,Mozgovoy:2008fd}, one can define other notions of Donaldson-Thomas invariants which are counted by $Z_{BPS}$. We will denote the corresponding Donaldson-Thomas partition functions by $\c\Zcal$. For specific orbifolds, \ie, finite subgroups of $SU(3)$, the definition is particularly simple. Namely, the Donaldson-Thomas invariants can be defined just as a sign weighted euler characteristic of a Hilbert scheme of points on the orbifold. In contrast, the definition of Donaldson-Thomas invariants at the non-commutative point is mathematically more sophisticated. Roughly said, one can see them as counting cyclic representations of a quiver with relations. 

At least for a certain class of models (backgrounds without compact divisor), there is a simple relation between the large volume and  the orbifold, respectively, non-commutative Donaldson-Thomas partition function \cite{Szendroi07,Young08,Aganagic:2009kf}. Namely, under a (model dependent) reparameterization of the non-commutative/orbifold partition function one has the relation
\beq\eqlabel{closedWC}
\c \Zcal(Q,q)=\Wcal \, \Zcal(Q,q)\,,
\eq
with 
\beq\eqlabel{Wfactor}
\Wcal=\t\Zcal(Q^{-1},q)\,,
\eq
where $\Zcal$ denotes the (large volume) Donaldson-Thomas partition function including degree $0$ contribution. (Recall that we have the factorization
\beq\eqlabel{deg0DTZfac}
\Zcal=\Zcal^0\,\t\Zcal\,,
\eq
with $\Zcal^0$ the degree $0$ part of the partition function.) We will refer to $\Wcal$ as wall crossing factor, since it arises in a BPS state counting perspective from crossing walls of marginal stability (separating the different chambers in K\"ahler moduli space). Especially, if we index the crossed walls by $i$ and cross an infinite number of walls while going from the non-commutative/orbifold to the large volume point, the wall crossing factor factorizes as $\Wcal=\prod_{i=1}^\infty \Wcal_i$, with $\Wcal_i$ the jump of the partition function under crossing the $i$th wall. This leads to the BPS state partition function in the $N$th chamber 
\beq\eqlabel{ZNBPS}
Z_{BPS}^{(N)}= \Zcal(Q,q)\, \prod_{i=N}^{\infty}\Wcal_i \,,
\eq
with $Z_{BPS}^{(1)}\equiv\c\Zcal$ (under reparameterization) and $Z_{BPS}^{(\infty)}\equiv\Zcal$. 

It is a natural question to ask how much of these results extend to the real case, \ie, hold in a similar fashion for specific orientifolds. As shown in \cite{Krefl:2009mw}, the real vertex possesses a combinatorial interpretation as well. This suggests that one should be able to define a real version of Donaldson-Thomas theory. The most naive approach to do so is to mimic the definition of real Gromov-Witten invariants, which can be seen as counting maps equivariant with respect to the orientifold projection. Hence, the most natural definition of real Donaldson-Thomas invariants is as counting equivariant sheaves. We will denote the corresponding partition function as $\Zcal^{real}$. In toric settings, the calculation of Gromov-Witten and Donaldson-Thomas invariants boils down to localization under the natural toric action, denoted by $\T$, on the respective moduli spaces. In the real case only the sublocus of the moduli spaces invariant under the subtorus $\T'\subset\T$ surviving the action of the orientifold projection contributes. This implies that 
\beq\eqlabel{realGWDTc}
\t\Zcal^{real}(Q,q)=\t Z^{real}(Q,e^{g_s}\rightarrow -q)\,,
\eq
should hold, constituting a real version of the Gromov-Witten/Donaldson-Thomas correspondence. The requirement that \req{realGWDTc} holds (which one may also see as simply defining the real Donaldson-Thomas partition function via the substitution $e^{g_s}\rightarrow -q$), translates to certain sign choices in the mathematical localization calculation of $\t\Zcal^{real}$. As we will discuss in section \ref{realDT}, the signs can be borrowed from the results of \cite{Krefl:2009mw}.

The Donaldson-Thomas formulation has certain advantages over the Gromov-Witten formulation in the real case. Essentially, while in the real Gromov-Witten case we have to make certain assumptions to be able to compute the partition function via localization on the moduli space of maps (\ie, that topological tapdole cancellation ensures that ordinary Hodge integrals are sufficient to compute the fixed-point contributions), in real Donaldson-Thomas theory in contrast there is a clear way to deal with the fixed points. Another advantage of the Donaldson-Thomas viewpoint is that it gives an intuitive way to deduce the degree $0$ contribution to $\Zcal^{real}$ (from which one can infer the constant map contribution to $Z^{real}$), as we will discuss in section \ref{d0contribution}. However, since Donaldson-Thomas theory is essentially a theory of signs (\cf, the localization calculation of \cite{MNOPI}), the prize one has to pay is that for the real version certain sign choices have to be made for consistency which are somehow more complex than the sign choices one has to perform in real Gromov-Witten localization calculations. 

Since the calculation of orbifold and non-commutative Donaldson-Thomas invariants similarly boils down to localization to $\T$ fixed-points, one can define in the same spirit a real version of these invariants by localizing with respect to the $\T'$ subtorus. However, as is already the case for the large volume real invariants, this localization might require certain additional sign choices which are not immediately clear how to derive from first principles. 

The real topological string partition function possesses a factorization into the purely closed and oriented topological partition function and a ``reduced" real partition function $Z'$ capturing the open and unoriented sector. Hence, $\Zcal^{real}$ possesses a similar factorization due to relation \req{realGWDTc}, \ie, 
\beq\eqlabel{Zcalsplit}
\Zcal^{real}= \Zcal^{1/2}\,  \Zcal'\,,
\eq
where it is understood that in $\Zcal^{1/2}$ the K\"ahler classes are identified according to the orientifold action on the second homology. Combined with the fact that the orientifold acts anti-holomorphic on the geometry, so holomorphic on the K\"ahler moduli space, it is immediate that a relation similar to \req{closedWC} should hold in the real case. That is, for every background for which \req{closedWC} with \req{Wfactor} holds, it is expected that correspondingly we have the relation
\beq\eqlabel{realWC}
\c\Zcal^{real}(Q,q)=\Wcal^{real}\,\Zcal^{real}(Q,q)\,,
\eq
with
\beq\eqlabel{realWCW}
\Wcal^{real}=\t\Zcal^{real}(Q^{-1},q)\,,
\eq
and with the same reparameterization of $\c\Zcal^{real}$ as for $\c\Zcal$. Especially, \req{closedWC} is valid for all models without compact divisors, and so we claim that \req{realWC} holds for the real orientifolds of these models. In turn, we can use \req{realWC} as guideline to deduce the correct sign insertions in the localization calculation of real orbifold, respectively real non-commutative invariants. We will discuss this mainly at hand of two examples. Namely, the  local $A_n$ singularity ($\C^2/\Z_n\times\C$ orbifold), to be discussed in section \ref{orbifoldsec}, and the non-commutative conifold, which we will discuss in section \ref{conisec}. 

Note that due to \req{Zcalsplit}, we can split the wall crossing factor as 
\beq\eqlabel{Wrealsplit}
\Wcal^{real}=\Wcal^{1/2} \, \Wcal'\,,
\eq
and isolate the wall crossing behavior of the reduced real partition function, \ie,
\beq
\c\Zcal'(Q,q)=\Wcal'\, \Zcal'(Q,q)\,.
\eq
Clearly, $\Wcal'=\t\Zcal'(Q^{-1},q)$. From \req{Zcalsplit} and the holomorphic action on the K\"ahler moduli space it is clear that the chamber structure is essentially unchanged (up to identification of chambers). Correspondingly, we expect as an analog of \req{ZNBPS}
\beq
{Z'}^{(N)}_{BPS}= \Zcal'(Q,q)\, \prod_{i>\lfloor N/2\rfloor }^{\infty}\Wcal_i' \,,
\eq
with ${Z'}^{(1)}_{BPS}\equiv \c\Zcal'$ (under reparameterization) and ${Z'}^{(\infty)}_{BPS}\equiv \c\Zcal'$. The partition function ${Z'}^{(N)}_{BPS}$ captures BPS states arising from the open and/or unoriented sector, \ie, D2-D0 bound states on curves which have boundaries and/or are unoriented, modulo the removal of some states as an imprint of topological tadpole cancellation. (The boundaries are provided by a D4 brane living in two space-time dimensions. It is the world-volume theory of this brane the BPS states live in). 

As is clear from the discussion above, a complete treatment of real (orientifold) wall crossing is beyond the scope of a single work. The aim of this note is rather to initiate the study of this topic by explaining what one should understand under real Donaldson-Thomas invariants at specific points in K\"ahler moduli space and giving evidence that the expected real wall crossing relation \req{realWC} indeed holds for certain models and sign choices. Hence, one may see this work roughly as a real analog of \cite{Szendroi07,Young08}. Especially, we are not going to discuss in detail the BPS state counting interpretation of the real partition functions or relation \req{realWC}, which we leave for subsequent work. 

The outline is as follows. In section \ref{realDT} we will discuss the large volume point in moduli space. Firstly, we briefly review the definition of ordinary Donaldson-Thomas invariants in section \ref{DTinvariants}. Then, we define real Donaldson-Thomas invariants following the tactic described above. This allows us to confirm and refine the constant map contribution to the real topological string, as we will describe in section \ref{d0contribution}. In section \ref{simplemodels}, we will discuss the real partition functions for a particular simple class of models for which one can express the partition functions completely in terms of (generalized) MacMohan functions. We will illustrate this at hand of the resolved conifold and the resolved local $A_n$ singularity in sections \ref{LVconifold} and \ref{LVC2Z2C}. 

In section \ref{orbifoldpoint} we will discuss the orbifold point. After giving the definition of real orbifold Donaldson-Thomas invariants in section \ref{orbidef}, we will briefly recall some basics about the transfer matrix approach in section \ref{transfermatrix}, which will be our main calculational tool to evaluate partition functions. In section \ref{orbifoldsec} we will discuss our main example for the orbifold point. Namely, the $\C^2/\Z_n\times\C$ orbifold. 

The non-commutative point will be discussed in section \ref{NCpoint}. The definition of real non-commutative Donaldson-Thomas invariants will be given in section \ref{defNCDT} and examples will be discussed in sections \ref{conisec} and \ref{c2z2Nsec}. We conclude in section \ref{concl}, where we as well offer a brief outline of possible followup directions of research.

In appendix \ref{appB} we give some useful expressions for generalized MacMohan functions. Appendix \ref{Schur} collects Schur function identities which are used in the main text. Some details of the main computations are redirected to appendix \ref{appC}.

\newpage

\section{Large volume point}
\label{realDT}

The purpose of this section is to give a definition of real Donaldson-Thomas invariants at the large volume point in K\"ahler moduli space. However, we will not attempt to give a fully rigorous treatment but rather aim at a qualitative construction to give a rough impression about what we mean by real Donaldson-Thomas partition functions/invariants in subsequent sections. The reader who is confident with taking \req{realGWDTc} as definition and is not interested in the degree $0$, respectively, constant map contribution, may directly proceed with section \ref{simplemodels}.

\subsection{Review of ordinary Donaldson-Thomas invariants}
\label{DTinvariants}
In order to set the stage and fix notation, let us first give a brief introduction to ordinary Donaldson-Thomas invariants of a local toric Calabi-Yau 3-fold $X$, mainly following \cite{MNOPI}. Let $I_n(X,\beta)$ denote the moduli space of ideal sheaves $\I$ on $X$ which satisfy
\beq\eqlabel{Icond1}
\chi(\O_Y)=n\,,
\eq
and
\beq\eqlabel{Icond2}
[Y]=\beta\in H_2(X,\Z)\,.
\eq
Here, $Y\subset X$ is the subscheme determined by $\I$ via the exact sequence
\beq
0\rightarrow \I\rightarrow \O_X\rightarrow \O_Y\rightarrow 0\,.
\eq
The moduli space $I_n(X,\beta)$ is isomorphic to the Hilbert scheme of curves of $X$, \ie, subschemes $Y\subset X$ with \req{Icond1} and \req{Icond2}. Especially, the degree $0$ moduli space $I_n(X,0)$ is simply the Hilbert scheme of $n$ points on $X$, which we will denote as $\Hilb^n(X)$. 

The Donaldson-Thomas invariant $d_{n,\beta}$ is defined via integration against the virtual fundamental class
\beq\eqlabel{DTinvdef}
d_{n,\beta}=\int_{\left[I_n(X,\beta)\right]^{vir}}1\,,
\eq
and the Donaldson-Thomas partition function as the formal sum
\beq\eqlabel{DTZ}
\Zcal=\sum_{\beta\in H_2(X,\Z)}\sum_{n\in \Z} d_{n,\beta}\, q^n Q^\beta\,,
\eq
where $q$ is a parameter and $Q^\beta:=Q_1^{d_1}\dots Q_l^{d_l}$ with $Q_i=e^{-t_i}$. We made a choice of basis $t_1,\dots, t_l$ of $H_2(X,\Z)$ such that any effective curve class $\beta$ is given by $\beta=\sum_i d_it_i$ with $d_i\geq 0$. Note that in the BPS state counting interpretation of $\req{DTZ}$ one can identify $n$ with the D0 charge and $\beta$ with the D2 charge of the state.

Factorizing $\Zcal$ as in \req{deg0DTZfac}, one can rigorously proof that the degree $0$ partition function is given by \cite{MNOPI,MNOPII}
\beq\eqlabel{DTdeg0Z}
\Zcal^0=M(1,-q)^{\chi(X)}\,,
\eq
where $M(1,q)$ is the MacMohan function (see \req{appBMh}) and $\chi(X)$ the euler characteristic of $X$. 

Since we take $X$ to be toric, we have a torus action on X, denoted as $\T$, which lifts to an action on $I_n(X,\beta)$ and hence reduces the integration in \req{DTinvdef} via virtual localization to a sum over fixed-points under $\T$. Let us denote the $\T$-fixed subspace of  $I_n(X,\beta)$ as  $I_n(X,\beta)^\T$. In order to evaluate \req{DTinvdef} we have to enumerate all $\I\in I_n(X,\beta)^\T$. 

For that, note that the to $\I$ associated subscheme $Y\subset X$ must be $\T$-fixed as well. Thus, $Y\subset X^\T$, where $X^\T$ denotes the invariant locus under the $\T$-action on $X$. The geometry of $X$ is full encoded in a toric fan, basically being a collection of $n$-cones. Each $n$-cone corresponds to a $3-n$ dimensional $\T$ invariant subspace of $X$. For $X$ being Calabi-Yau, one can project the toric fan to a two dimensional graph, called the toric diagram of $X$. Taking the graph dual, we obtain a linear trivalent graph $\Gamma$ also known as $(p,q)$-web diagram. Note that the $(p,q)$-web precisely corresponds to the part of $X^\T$ with complex dimension smaller two, \ie, the vertices of the web correspond to the fixed points while the edges to the fixed curves under $\T$. Let us denote the set of vertices of the web by $V$, the vertices by $v_\alpha$ and an edge connecting the vertices $v_\alpha$ and $v_\beta$ by $e_{\alpha\beta}$ (where is at most one edge between two vertices). The set of all edges $e_{\alpha\beta}$ will be referred to as $E$.

Around each $v_\alpha$ where is a canonical $\T$-fixed open chart $U_\alpha$ on which $\I$ is given by a monomial ideal $I_\alpha$, \ie,
\beq\eqlabel{monideal}
I_\alpha=\I|_{U_\alpha}\subset \C[x_1,x_2,x_3]\,,
\eq 
where $\C[x_1,x_2,x_3]$ denotes the polynomial ring over $\C^3$. Since a monomial ideal is equivalent to a (not necessarily finite) 3d partition, we can identify each $\I|_{U_\alpha}$ with a 3d partition $\pi_\alpha$, \ie, we take
\beq
\pi_\alpha=\left\{(i,j,k): x_1^ix_2^jx_3^k \notin I_\alpha \right\}.
\eq
Recall that a 3d partition $\pi$ is a set of boxes $(i,j,k)\in \Z^3_{\geq 0}$ such that if any of the boxes with coordinates $(i+1,j,k)$, $(i,j+1,k)$, $(i,j,k+1)$ are in $\pi$, so is the box with coordinate $(i,j,k)$. We define the volume of the partition $|\pi|$ to be the number of boxes it contains, and the renormalized volume $|\pi|'$ as 
\beq
|\pi_\alpha|'=|\pi_\alpha \cap[0,\dots,N^3] |-(N+1)\sum_{i=1}^3|\lambda_{\alpha i}|\,,
\eq
where $N\gg 0$ is a cutoff and $\lambda_{\alpha i}$ the asymptotics of $\pi_\alpha$ along the $i$th axis, \ie, $\lambda_{\alpha i}$ is a 2d partition (Young diagram).

It remains to determine $\I|_{U_\alpha\cap U_\beta}$. But this is exactly what the set of edges of the web tell us. Namely, each edge $e_{\alpha\beta}$ can be associated with an intersection $U_\alpha\cap U_\beta$. Since an edge $e_{\alpha\beta}$ corresponds to a $\P^1$, the local geometry encoded by the edge is given by a local curve, that is $\Ncal_{\alpha\beta}\rightarrow \P^1$, where $\Ncal_{\alpha\beta}$ denotes the normal bundle of the $\P^1$ in $X$, \ie, 
\beq
\Ncal_{\alpha\beta}=\O(m_{\alpha\beta})\oplus \O(-2-m_{\alpha\beta})\,.
\eq
Let us take the coordinate $x_1$ of the chart $U_\alpha$ to be a coordinate of one of the two canonical patches of the $\P^1$, while the coordinates $x_2$ and $x_3$ are taken to be on the normal bundle. Similarly, for $U_\beta$ we take $x_1'=1/x_1$ to be a coordinate on the other patch of the $\P^1$ while $x_2'=x_2$ and $x_3'=x_3$ are coordinates on the bundle. The transition function between the patches $U_\alpha$ and $U_\beta$ reads
\beq
(x_1,x_2,x_3)\rightarrow (x_1^{-1}, x_1^{-m_{\alpha\beta}} x_2 ,x_1^{2+m_{\alpha\beta}} x_3)\,.
\eq
In particular, along the $x_1$ axis corresponding to the coordinate of the $\P^1$, the asymptotic of $\pi_\alpha$ and $\pi_\beta$ (with $\pi_i$ associated to $U_i$) is the 2d partition $\lambda_{\alpha\beta}$,
\beq\eqlabel{edgepartition}
\lambda_{\alpha\beta}=\left\{(j,k): x_2^j x_3^k\notin I_{\alpha\beta} \right\},
\eq
with
\beq
I_{\alpha\beta}=\I|_{U_\alpha\cap U_\beta}\subset \C[x_1^{\pm 1},x_2,x_3] \,.
\eq

We conclude that the set of $\I$ (the $\T$ fixed points) is encoded via the following combinatorial data
\beq\eqlabel{Pidef}
\Pi_\I=\{\pi_\alpha;\lambda_{\alpha\beta_1},\lambda_{\alpha\beta_2},\lambda_{\alpha\beta_3}\}\,.
\eq 
That is, the set of 3d partitions $\pi_\alpha$ with boundary condition along the axis given by the three 2d partitions $\lambda_{\alpha\beta_i}$ (recall that each vertex of the web is trivalent).

Virtual localization tells us that \cite{MNOPI}
\beq\eqlabel{DTlocresult}
d_{n,\beta}:=\int_{\left[I_n(X,\beta)\right]^{vir}}1=\sum_{\Pi_\I} (-1)^{\sum_\alpha |\pi_\alpha|'+\sum_{\alpha<\beta}f(m_{\alpha\beta})+\sum_{\alpha<\beta} m_{\alpha\beta}|\lambda_{\alpha\beta}|}\,,
\eq
with
\beq\eqlabel{DTfrel}
f(m_{\alpha\beta})=\sum_{(i,j)\in \lambda_{\alpha\beta}}\left(m_{\alpha\beta}(j-i)+2(j-1)+1\right)\,.
\eq
 The explicit expression of the Donaldson-Thomas invariant given above allows us to rewrite the partition function \req{DTZ} as
\beq\eqlabel{DTZfinal}
\Zcal=\sum_{\Pi_\I}(-1)^{\sum_{\alpha,\beta} m_{\alpha\beta}|\lambda_{\alpha\beta}|}  (-q)^{\sum_\alpha |\pi_\alpha|'+\sum_{\alpha<\beta}f(m_{\alpha\beta})} \prod_{\alpha<\beta}Q_{\alpha\beta}^{|\lambda_{\alpha\beta}|}\,.
\eq
Using the relations
\beq\eqlabel{kappadef}
\kappa(\lambda)=2\sum_{(i,j)\in\lambda}(j-i)\,,
\eq
and
\beq
||\lambda||^2=2|\lambda|+2\left(
\begin{matrix}
\lambda\\
2
\end{matrix}
\right)\,,
\eq
we can rewrite \req{DTfrel} as
\beq
f(m_{\alpha\beta})=\frac{(m_{\alpha\beta}+1)}{2}\kappa(\lambda_{\alpha\beta})+\frac{||\lambda_{\alpha\beta}||^2+||\lambda_{\alpha\beta}^t||^2}{2}\,.
\eq
Setting $n_{\alpha\beta}=m_{\alpha\beta}+1$,
we deduce
\beq
\Zcal=\sum_{\Pi_\I}(-1)^{\sum_{\alpha<\beta} (n_{\alpha\beta}+1 )|\lambda_{\alpha\beta}|}  (-q)^{\sum_\alpha |\pi_\alpha|'+\sum_{\alpha<\beta}\frac{n_{\alpha\beta}\kappa(\lambda_{\alpha\beta})+||\lambda_{\alpha\beta}||^2+||\lambda_{\alpha\beta}^t||^2 }{2} } \prod_{\alpha<\beta}Q_{\alpha\beta}^{|\lambda_{\alpha\beta}|}\,.
\eq

Using the relation between the topological vertex and a 3d partition $\pi_\alpha$ with boundary partitions $(\lambda_{\alpha \beta_1},\lambda_{\alpha \beta_2},\lambda_{\alpha \beta_3})$ \cite{Okounkov:2003sp}
\beq\eqlabel{vertexrelation}
C_{\lambda_{\alpha \beta_1}\lambda_{\alpha \beta_2}\lambda_{\alpha \beta_3}}=M(1,q)^{-1}q^{\frac{1}{2}(||\lambda^t_{\alpha \beta_1}||^2+||\lambda^t_{\alpha \beta_2}||^2+||\lambda^t_{\alpha \beta_3}||^2)}\sum_{\pi_\alpha\in\Pcal'} q^{|\pi_\alpha|'}\,,
\eq
where $\Pcal'$ denotes the set of 3d partitions with boundaries $(\lambda_{\alpha \beta_1},\lambda_{\alpha \beta_2},\lambda_{\alpha \beta_3})$, we immediately deduce that 
\beq\eqlabel{GWDTcor}
\Zcal=M(1,q)^{\chi}\t Z(-e^{g_s})\,.
\eq
Equation \req{GWDTcor} is  the celebrated Gromov-Witten/Donaldson-Thomas correspondence of \cite{MNOPI} (see also \cite{Iqbal:2003ds}). Note that we observe from \req{GWDTcor} that $\Zcal^0=\left(Z^{0}\right)^{2}$, where $Z^0$ denotes the constant map contribution to the topological string partition function, in agreement with \req{DTdeg0Z}.

\subsection{Real Donaldson-Thomas invariants}
\label{realDTinvariants}
Let us now discuss the real case. For that, let us consider the pair $(X,\inv$) with
\beq\eqlabel{invdef}
\inv:X\rightarrow X\,,
\eq
an anti-holomorphic involution acting on the coordinates $x_i$ of $X$ as $x_i\rightarrow M_{ij} \cc x_j$ with $M_{ij}$ a phase. In general, the action of $\inv$ leaves only a subtorus $\T'\subset\T$ of the original torus action of $X$ intact. It is this subtorus which is used to define real Gromov-Witten invariants via localization on the moduli space of maps \cite{Walcher:2006rs,Walcher:2007qp,Krefl:2009md} (combined with a choice of signs). As outlined in the introduction, we define in a similar spirit real Donaldson-Thomas invariants via localizing \req{DTinvdef} with respect to the same subtorus and choosing signs appropriately. The choice of signs is motivated by the expectation that the Gromov-Witten/Donaldson-Thomas correspondence \req{GWDTcor} should still hold in the real case, \ie,  
we arrange the signs such that
\beq
\t\Zcal^{real}(Q,q)=\t Z^{real}(Q,e^{g_s}\rightarrow -q)\,,
\eq
holds. One might see this as simply defining $\t \Zcal^{real}$ as the partition function of the real topological string $\t Z^{real}$ under the substitution $e^{g_s}\rightarrow -q$. 

In the remainder of this section we will show that this is indeed well defined, \ie, that there exists a choice of signs such that we can recover $Z^{real}(Q,e^{g_s}\rightarrow -q)$ from localizing \req{DTinvdef} under the action of the torus $\T'$. Fortunately, we can borrow most of the technicalities from the real topological vertex formalism of \cite{Krefl:2009mw}. 

Note first that the $\inv$ action induces a split of $H_2(X,\Z)$ into 
\beq\eqlabel{H2grading}
H_2(X,\Z)=H^a_2(X,\Z)\oplus H^b_2(X,\Z)\,,
\eq
where $a$ refers to the subgroup of $H_2(X,\Z)$ formed by the invariant elements under $\inv$ and $b$ to the subgroup formed by the non-invariant elements. Let us denote a basis of $H^a_2(X,\Z)$ by $t^a_1,\dots,t^a_l$ and a basis of $H^b_2(X,\Z)$ by $t^b_1,\dots,t^b_k$. Note that $\dim H^b_2(X,\Z)\equiv 0\,{\rm mod}\, 2$. The effective curve class $\beta$ splits accordingly, \ie, $\beta=\beta_a+\beta_b=\sum_i d^a_i t_i^a +\sum_i d^a_i t_i^b$. The quotient can be implemented by replacing $Q_i\rightarrow Q_i^{1/2}$ (clearly, the curves fixed under the involution have half the volume of the corresponding covering space curve, whereas the non-invariant curves pair up with their mirror image and cancel the factor of $1/2$). By abuse of notation, we will sometimes  refer to $\beta_b$ after identification of moduli and dividing by $1/2$ just as $\beta_b$. 

The explicit splitting \req{H2grading} can be read of from the action of $\inv$ on the web diagram. If we denote the set of edges as $E$ and set of vertices as $V$, $\inv$ acts on the web diagram as $\inv:V\rightarrow V$ and $\inv:E\rightarrow E$. Both, the set of vertices and the set of edges splits into an invariant and non-invariant set under $\inv$, \ie,
\beq\eqlabel{Vsplit}
\begin{split}
V&=V_a\cup V_b\,.\\
E&=E_a\cup E_b\,.\\
\end{split}
\eq
Since the edges are associated with elements of $H_2(X,\Z)$, we can infer the splitting \req{H2grading} from the splitting of $E$.

\begin{figure}
\begin{center}
\psfrag{x}[cc][][0.7]{$x_1$}
\psfrag{z}[cc][][0.7]{$x_2$}
\psfrag{y}[cc][][0.7]{$x_3$}
\includegraphics[scale=0.3]{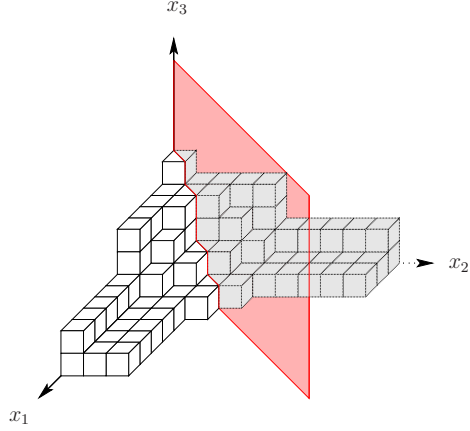}
\end{center}
\caption{Illustration of a symmetric 3d partition $\pi\subset\Scal$. Note that the volume of the partition is halved in the quotient.}
\label{3dsympart}
\end{figure}

Let us consider first the vertices. Note that nothing changes for the set $V_b$, \ie, we can still associate a 3d partition $\pi_\beta$ to each $v\in V_b$ because locally on these charts the $\inv$ action keeps the full original $\T$ intact. For the set $V_a$ in contrast, the orientifold acts locally as (we consider only line reflections and point-reflections sitting on edges of the web diagram, \cf, \cite{Krefl:2009mw})
\beq
\inv:(x_1,x_2,x_3)\rightarrow (\cc x_2,\cc x_1,\cc x_3)\,,
\eq
and preserves only a $\T'=\C^*$ of the original $\T$ action. We conclude that the $\inv$ invariant $I_\alpha$ are generated by monomials $x_1^i x_2^j x_3^k$ with $(i,j,k)\in\pi_\alpha\in\Scal$, where $\Scal\subset\Pcal$ denotes the subset of symmetric plane partitions, that is, if $(i,j,k)\in\pi_\alpha$ so is $(j,i,k)\in\pi_\alpha$ (see figure \ref{3dsympart}).
Thus, we associate the ideal on a fixed chart with a symmetric plane partition. Especially, the partition $\lambda_{\alpha\beta_3}$ asymptotic on the $x_3$ axis needs to be self-conjugate, as is clear from figure \ref{3dsympart}. However, one should note that the volume of the partition is halved in the quotient. Hence, for a single (fixed) vertex we obtain from \req{DTZfinal} the partition function (\cf, \req{realMacMohanstandard})
\beq\eqlabel{Zdeg0}
\Zcal^{real}_\pm=\sum_{\pi\in\Scal }\, (\mp 1)^{\tr(\pi)} \,(-q)^{|\pi|/2}=M^{real}_\pm(1,-q)\,,
\eq
where $\tr(\pi)$ denotes the number of boxes on the diagonal slice of $\pi$, \ie, the number of boxes on the reflection symmetry plane illustrated in figure \ref{3dsympart}  (since $\pi$ is symmetric, $(\pm 1)^{|\pi|}=(\pm 1)^{\tr(\pi)}$). The origin of the sign can be seen in taking a squareroot of $(-q)$. 

Let us now go on to the edges. Clearly, for the edges belonging to $E_b$ nothing changes. However, for the edges in $E_a$ we have to distinguish between three cases. Either the orientifold acts on the edge $e_{\alpha\beta}\in E_a$ as a point-reflection, as a line reflection along the edge or as a line reflection orthogonal to the edge \cite{Krefl:2009mw}. More explicitly, the different orientifolds of the local $\Ncal_{\alpha\beta}\rightarrow \P^1$ geometry represented by the edge $e_{\alpha\beta}$ act on the coordinates of the geometry as follows
 \beq\eqlabel{localgeooaction}
 \begin{split}
 \sigma_{\emptyset}&: (x_1,x_2,x_3)\rightarrow (-\cc x_1^{-1},\cc x_3,-\cc x_2)\,,\\
 \sigma_{|}&: (x_1,x_2,x_3)\rightarrow (\cc x_1^{-1},\cc x_3,\cc x_2)\,,\\
 \sigma_{\bot}&: (x_1,x_2,x_3)\rightarrow (\cc x_1^{-1},\cc x_2,\cc x_3)\,.
 \end{split}
 \eq
Especially, $\sigma_\emptyset$ and $\sigma_\bot$ act on the bundle $\Ncal_{\alpha\beta}=\O(m_{\alpha\beta})\oplus \O(-2-m_{\alpha\beta})$ via identification of the two summands. Thus, we have that necessarily $m_{\alpha\beta}=-1$ for these cases and these two actions are only consistent if the local geometry is a resolved conifold. The first action acts free, while the second possesses a fixed point locus being topologically $S^1\times \R^2$ (\cf, \cite{Hori:2005bk}). In contrast, $\sigma_\bot$ maps each summand of the bundle to itself, hence $m_{\alpha\beta}$ is unrestricted. The topology of the fixed-point locus is as for $\sigma_|$.

\begin{figure}
\begin{center}
\includegraphics[scale=0.4]{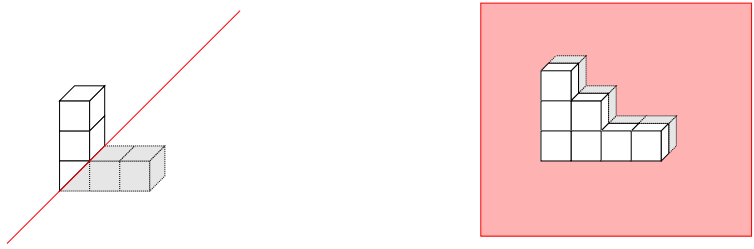}
\end{center}
\caption{Left: Illustration of the action of $\sigma_\emptyset$ and $\sigma_|$ on $\lambda_{\alpha\beta}$. Right: The action of $\sigma_\bot$. Especially, the volume of $\lambda_{\alpha\beta}$ is always halved in the quotient.}
\label{2dpartitions}
\end{figure}

From the actions \req{localgeooaction} and \req{edgepartition} we deduce that for $\sigma_\emptyset$ and $\sigma_|$ the 2d partition $\lambda_{\alpha\beta}$ needs to be self-conjugate, \ie, $\lambda_{\alpha\beta}=\lambda_{\alpha\beta}^t$, while for $\sigma_\bot$ the partition $\lambda_{\alpha\beta}$ is unrestricted, see also figure \ref{2dpartitions}. 
In the quotient the volume of the partition is halved for both cases. We therefore weight each box by $1/2$. Therefore, \req{DTfrel} gains a factor of $1/2$ for the edges in $E_a$. This leads us to the real Donaldson-Thomas invariants 
\beq
D_{n,\beta_a,\beta_b}=\sum_{\Pi^a_{\I},\Pi^b_{\I}} (-1)^{\left[\sum_{\alpha<\beta}m_{\alpha\beta}\sigma(\lambda_{\alpha\beta})+\sum_\alpha |\pi_\alpha|'+\sum_{\alpha<\beta}f(m_{\alpha\beta})+\sum_{\alpha<\beta} m_{\alpha\beta}|\lambda_{\alpha\beta}|\right]/2}\prod_{\alpha=1}^{|\Pi^a_\I|}\sigma_\alpha^{\tr(\pi_\alpha)}  \,,
\eq
where $\Pi^a_{\I}=\left\{\pi_\alpha\in\Scal; \lambda_{\alpha\beta_1}=\lambda_{\alpha\beta_2}, \lambda_{\alpha\beta_3}=\lambda_{\alpha\beta_3}^t\right\}$ with $\alpha$ running over the elements of $V^a$, $\Pi^b_{\I}$ as in \req{Pidef} with index running over $V^b$, and $\sigma_\alpha$ a sign which corresponds in the real topological vertex formalism to the choice between twisted versus non-twisted topological vertex associated to $\pi_\alpha\in\Scal$. Furthermore, we performed an ad hoc sign insertion of 
\beq
\sigma(\lambda_{\alpha\beta})=\left\{
\begin{matrix}
0 &&& {\rm for} & e_{\alpha\beta}\in E^b&&\\
r(\lambda_{\alpha\beta}) &&& {\rm for} & e_{\alpha\beta}\in E^a &{\rm and}&\sigma_\emptyset, \sigma_{|}\\
c(\lambda_{\alpha\beta}) &{\rm or}& c(\lambda_{\alpha\beta}^t) & {\rm for} & e_{\alpha\beta}\in E^a&{\rm and}&\sigma_\bot\\
\end{matrix}
\right\}\,,
\eq
where $r(\lambda_{\alpha\beta})$ denotes the rank of the Young diagram, \ie, the number of boxes on the diagonal, and $c(\lambda_{\alpha\beta})$ the number of columns of odd height of $\lambda_{\alpha\beta}$. Note that the choices of signs $\sigma_\alpha$ and $c(\lambda_{\alpha\beta})$ versus $c(\lambda_{\alpha\beta}^t)$ is correlated, as discussed in detail in \cite{Krefl:2009mw}. 

Roughly, the real Donaldson-Thomas invariant can be seen as arising from the integration over the moduli space of ideal sheaves on $X$ which are invariant under the action of the (anti-holomorphic) involution $\sigma$.

We assemble the real invariants into the partition function
\beq
\Zcal^{real}= \sum_{\beta \in H_2(X,\Z)}\sum_{n\in\Z} \sigma^{\beta_a} D_{n,\beta_a,\beta_b} \,q^{n/2} Q^{\beta_a/2} Q^{\beta_b}\,,
\eq
with $\sigma^{\beta_a}=\prod_{i=1}^{\dim H_2^a(X,\Z)} \sigma_i^{d^a_i}$ and $\sigma_i$ a sign. From a physical point of view, one should note that now the D0 and D2 charge can take fractional, \ie, half-integer, values.

It is immediate from the construction that 
\beq\eqlabel{realDTGWcor}
\t \Zcal^{real}(Q,q)=\t Z^{real}(Q,-e^{g_s})\,,
\eq
as we wanted to observe for our sign insertions. However, in order to explicitly show this relation in full generality, one would need the real analog of \req{vertexrelation}, which so far has been given only for the vertex with trivial partition along the $x_3$ axis \cite{Krefl:2009mw}.

\subsection{Degree $0$ contribution}
\label{d0contribution}

After the previous discussions, it is easy to infer how one can deduce the degree $0$ contribution. In terms of the web diagram, one can see the degree $0$ contribution simply as removing the edges $e_{\alpha\beta}$ such that the web diagram is a collection of vertices.  Correspondingly, in this case $\Pi_\I$ is a disjoint union of 3d partitions, \ie, $\Pi_\I=\{\pi_\alpha;\emptypar,\emptypar,\emptypar\}$. We deduce (via simply setting $|\lambda_{\alpha\beta}|=0$ in \req{DTZfinal} and letting $N\rightarrow\infty$)
\beq
\Zcal^0=\prod_\alpha\sum_{\pi_\alpha\in\Pcal}(-q)^{|\pi_\alpha|}= M(1,-q)^{|v|},
\eq
where $|v|$ is the number of vertices of the web diagram of $X$. Recall that the web is graph-dual to the toric diagram, so each vertex corresponds to a 3-cone. Since the number of 3-cones is simply $\chi(X)$ we obtain \req{DTdeg0Z}.

The degree $0$ contribution in the real case can be obtained in a similar fashion. We split the set of vertices as in \req{Vsplit} of section \ref{realDTinvariants}. To each element of $V_b$ we associate a MacMohan function $M(1,q)$ and divide by two, since we are interested in the quotient. For each element of $V_a$ instead, we associate a real MacMohan $M^{real}_\pm(1,q)$. 
We conclude that
\beq\eqlabel{realdeg0Z}
{\Zcal^0_\sigma}^{real}=\prod_{\alpha,\beta} \sum_{\pi_\alpha\in\Pcal} (-q)^{|\pi_\alpha|} \sum_{\pi_\beta\in\Scal}\sigma_\beta^{\tr(\pi_\beta)} (-q)^{|\pi_\beta|/2}=M(1,-q)^{|V_b|/2}\prod_{i=1}^{|V_a|}M^{real}_{\sigma_i}(1,-q) \,,
\eq
where $\alpha$ runs over $V_a$, $\beta$ over $V_b$ and $|V_a|$ denotes the number of vertices fixed under the action of $\inv$ and $|V_b|$ the number of non-fixed vertices. Clearly, $|V_a|+|V_b|=|V|$. Further, we introduced a sign $\sigma=(\sigma_1,\dots,\sigma_{|V_a|})$, where the possible choices of $\sigma_i$ are globally correlated in a similar fashion as the choice between real versus twisted real vertex in the real vertex formalism of \cite{Krefl:2009mw}.

Let us define $\chi_a=|V_a|$, $\chi_b=|V_b|/2$, take the squareroot of \req{realdeg0Z}, perform the substitution $q\rightarrow -e^{g_s}$  and for simplicity assume that all $\sigma_i$ are equal, \ie, $\sigma_i=\pm 1$. Under these conditions we recover the qualitative form of the real topological string constant map contribution conjectured in \cite{Krefl:2009mw}, \ie,
\beq
Z^{real}_\pm\sim_{t_k\rightarrow \infty} M^{real}_\pm(1,q)^{\frac{\chi_a}{2}}M(1,q)^{\frac{\chi_b}{2}}\,,
\eq
with 
\beq\eqlabel{chisplit}
\chi_a+2\chi_b=\chi(X)\,.
\eq
However, one should note that we now know the value of $\chi_a$ for a local toric Calabi-Yau 3-fold $X$ and the global constraints on the signs $\sigma_i$.

Finally, from the factorization \req{Zcalsplit} and invoking \req{appBreducedrealMac}, we deduce the degree $0$ contribution to the reduced real Donaldson-Thomas partition function (defined via equation \req{Zcalsplit}) to be
\beq
{{\Zcal^0_\sigma}'}^{real} = \prod_{i=1}^{\chi_a}\prod_{n=1}^\infty \left(\frac{1-\sigma_i (-q)^{n-1/2}}{1+\sigma_i (-q)^{n-1/2}}\right)^{1/2}\,.
\eq
A similar result holds for the constant map contribution to the reduced real topological string partition function $Z'$.

\section{A simple class of backgrounds}
\label{simplemodels}
In this section, we will derive the large volume real partition functions of the two main examples we will consider at different points in K\"ahler moduli space in the remaining sections.  Namely, the conifold and the $\C^2/\Z_n\times\C$ orbifold. These models fall into a class of models for which it is particular easy to qualitatively derive the real partition functions, as we will explain below.

\subsection{Large volume partition functions}
An intriguing feature of the closed topological string on some specific class of local backgrounds is that one can express the topological partition function as a product of generalized MacMohan functions $M(x,q)$, \ie,
\beq\eqlabel{Zgeneral}
\t Z=\prod_a M(Q_1^{f_1(a)}\dots Q_n^{f_n(a)},q)^{g(a)}\,,
\eq
where $q=e^{g_s}$ with $g_s$ the string coupling, $Q_i=e^{-t_i}$ with $t_i$ the K\"ahler moduli of the background (we assume there are $n$), $a\in \N$ a model dependent parameter and $f_i(a)$ and $g(a)$ some model dependent functions $f_i: \N\rightarrow \N$, respectively $g:\N\rightarrow\Z$. The simplest example of a background for which \req{Zgeneral} holds is the resolved conifold, for which we just have $a=f_1(a)=g(a)=1$. Other well-known examples are the topological vertex geometry \cite{Karp:2005vq} and the resolution of a orbifold $\C^2/\Z_n\times\C$ \cite{Young08,Bryan08}.

The backgrounds $X$ for which \req{Zgeneral} holds can be deduced as follows. Recall that one can express the topological string partition function in terms of Gopakumar-Vafa invariants $n^{(g)}_{\vec d}$ ($\vec d$ labels the K\"ahler class, \ie, $\vec d=(d_1,d_2,\dots, d_n)$) as \cite{Katz:1999xq,Klemm:2004km} 
\beq\eqlabel{ZclosedInf}
\t Z=\prod_{\vec d, j>0}(1-Q^{\vec{d}} \,q^j)^{j n_{\vec d}^{(0)}} \prod_{\vec d, g>0}\prod_{l=0}^{2g-2}(1-Q^{\vec{d}}\, q^{g-l-1})^{(-1)^{g+l}  
\left(\topa{2g-2}{l}\right)n^{(g)}_{\vec d}}\,,
\eq
where $Q^{\vec d}=Q_1^{d_1}Q_2^{d_2}\dots Q_n^{d_n}$.

If all higher genus Gopakumar-Vafa invariants vanish, this simply becomes (\cf, \req{appBMh})
\beq\eqlabel{Zpro}
\t Z=\prod_{\vec d}M(Q^{\vec d},q)^{-n_{\vec d}^{(0)}}\,.
\eq
Hence, \req{Zgeneral} holds for all models with $n^{(g>0)}_{\vec d}=0$.  In detail, the parameter $a$ in \req{Zgeneral} can be identified with $\vec d$, $f_i(a)$ with $d_i$ and $g(a)$ with $-n^{(0)}_{\vec d}$. 

\begin{figure}
\psfrag{1}[cc][][0.7]{$t_1$}
\psfrag{2}[cc][][0.7]{$t_2$}
\psfrag{3}[cc][][0.7]{$t_3$}
\psfrag{a}[cc][][0.7]{$R_1$}
\psfrag{b}[cc][][0.7]{$R_2$}
\psfrag{c}[cc][][0.7]{$R_3$}
\begin{center}
\includegraphics[scale=0.6]{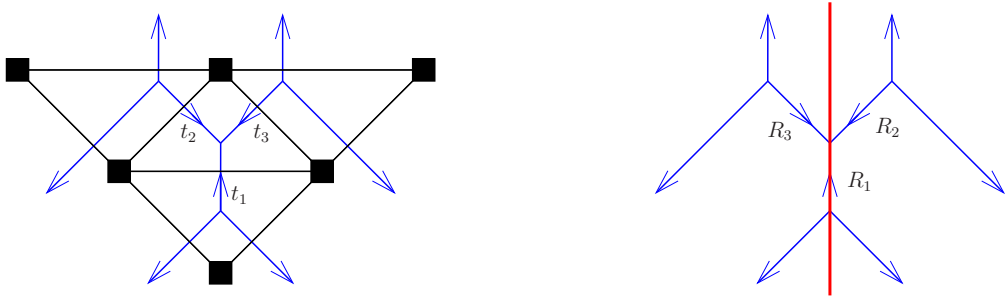}
\end{center}
\caption{Left: Toric diagram and $(p,q)$-web of the topological vertex geometry, \ie, a resolution of $\C^2/(\Z_2\times\Z_2)\times\C$. Right: The action of the anti-holomorphic involution on the web diagram (reflection along the red line).}
\label{topvertexgeofig}
\end{figure}

Let us illustrate \req{Zpro} at hand of the topological vertex geometry. The topological vertex geometry corresponds to the resolution of the $\C^2/(\Z_2\times\Z_2)\times\C$ singularity shown in figure \ref{topvertexgeofig} and possesses three K\"ahler parameter $t_i$. 
The non-vanishing Gopakumar-Vafa invariants are \cite{Karp:2005vq} 
\beq\eqlabel{GVtopvertex}
\begin{split}
n^{(0)}_{1,1,0}&=n^{(0)}_{1,0,1}=n^{(0)}_{0,1,1}=-1\,,\\
n^{(0)}_{1,0,0}&=n^{(0)}_{0,1,0}=n^{(0)}_{0,0,1}=n^{(0)}_{1,1,1}=1\,.
\end{split}
\eq
Thus, we have
\beq
\t Z=\frac{M(Q_1Q_2,q)M(Q_1Q_3,q)M(Q_2Q_3,q)}{M(Q_1,q)M(Q_2,q)M(Q_3,q)M(Q_1Q_2Q_3,q)}\,.
\eq

It is an interesting question to ask if a similar simple structure as \req{Zpro} holds in the real case. In order to answer this question, note that the real topological string partition function possesses an integer expansion into real Gopakumar-Vafa invariants $N_{\vec d_a,\vec d_b}^{(\chi)}$ \cite{Walcher:2007qp}, which is very similar to the original closed topological string Gopakumar-Vafa expansion \cite{Gopakumar:1998ii,Gopakumar:1998jq}. Namely, the (reduced) real free energy can be expanded as
\beq\eqlabel{realexpansion}
\Gcal'_\sigma=\sum_{\substack{\chi\geq -1\\ \vec d_a,\vec d_b\\k\,{\rm odd}}}\sigma^{k}_{ \vec d_a} {N_{\vec d_a,\vec d_b}^{(\chi)}}\frac{1}{k}\left(2\sinh\left(\frac{k \lambda }{2}\right) \right)^{\chi} Q_{\vec d_a}^{k  /2}Q_{\vec d_b}^{k }\,,
\eq
where $\vec d_a$ denotes the part of the K\"ahler classes associated to the moduli invariant under the orientifold action, and $\vec d_b$ to the part which is not invariant (here, $\vec d_b$ is to be understood in the quotient, \ie, after identification of moduli). The summation runs over the euler number $\chi$ which is related to the genus $\h g$ of the covering space curve via $\chi=\h g-1$ and we inserted an additional sign degree of freedom $\sigma_{ \vec d_a}$ given by
\beq
\sigma_{ \vec d_a}=\prod_{i=1}^{|\vec d_a|} \sigma_i^{d_i}\,,
\eq
with $\sigma=(\sigma_1,\dots,\sigma_{|\vec d_a|})$ a set of arbitrary signs. The origin of these signs is the squareroot of $Q^k_{\vec d_a}$ and one may absorb them into the real Gopakumar-Vafa invariants.

Using the fundamental relation $\t Z'_\sigma=\exp \Gcal'_{\sigma}$, it is straight-forward to express the reduced real topological partition function $\t Z'_\sigma$ as an infinite product, similar as \req{ZclosedInf}. One obtains
\beq\eqlabel{ZrealInf}
\begin{split}
\t Z'_\sigma=&\prod_{\substack{\vec d_a,\vec d_b\\n>0}}\left(\frac{1-\sigma_{ \vec d_a} Q_{\vec d_a}^{1/2} Q_{\vec d_b}q^{n-1/2}}{1+\sigma_{ \vec d_a} Q_{\vec d_a}^{1/2} Q_{\vec d_b}q^{n-1/2}}\right)^{-N^{(-1)}_{\vec d_a,\vec d_b}/2}\\
&\times\prod_{\substack{\chi\geq 0\\\vec d_a,\vec d_b}}\prod_{n=0}^\chi\left(\frac{1-\sigma_{ \vec d_a} Q_{\vec d_a}^{1/2} Q_{\vec d_b}q^{\chi/2-n}}{1+\sigma_{ \vec d_a} Q_{\vec d_a}^{1/2} Q_{\vec d_b}q^{\chi/2-n}} \right)^{-N^{(\chi)}_{\vec d_a,\vec d_b} (-1)^n\left(\topa{\chi}{n}\right)/2}\,.
\end{split}
\eq
As for the ordinary topological string partition function, we can express the $\chi=-1$ part of \req{ZrealInf} in terms of generalized MacMohan functions such that for models with $N^{(\chi\geq 0)}_{\vec d_a,\vec d_b}=0$ we simply have 
\beq\eqlabel{realredZpro}
\t Z'_\sigma=\prod_{\vec d_a,\vec d_b,n>0}\left(\frac{1-\sigma_{ \vec d_a} Q_{\vec d_a}^{1/2} Q_{\vec d_b}q^{n-1/2}}{1+\sigma_{ \vec d_a} Q_{\vec d_a}^{1/2} Q_{\vec d_b}q^{n-1/2}}\right)^{-N^{(-1)}_{\vec d_a,\vec d_b}/2}=\prod_{\vec d_a,\vec d_b}M'_{\sigma_{ \vec d_a}}(Q_{\vec d_a}Q^2_{\vec d_b},q)^{-N^{(-1)}_{\vec d_a,\vec d_b}}\,,
\eq 
where $M'_{\sigma_{ \vec d_a}}(x,q)$ is the generalized reduced real MacMohan function defined in \req{appBreducedrealMac}. 

Since we have the relation
\beq\eqlabel{reducedZtop}
\t Z':=\frac{\t Z^{real}}{\t Z^{1/2}}\,,
\eq
we infer from \req{Zpro} and \req{realredZpro}
\beq\eqlabel{realZpro}
\begin{split}
\t Z^{real}_\sigma=& \prod_{\vec d} M(Q_{\vec d_a}Q_{\vec d_b}Q_{\vec d_c},q)^{-n^{(0)}_{\vec d}/2} \\ &\times \prod_{\vec d' }M(Q_{\vec d_a}Q^2_{\vec d_b},q)^{\frac{-n^{(0)}_{\vec d'}+N^{(-1)}_{\vec d_a,\vec d_b}}{2}}M^{real}_{\sigma_{\vec d_a}}(Q_{\vec d_a}Q^2_{\vec d_b},q)^{-N^{(-1)}_{\vec d_a,\vec d_b}}\,,
\end{split}
\eq
where $\vec d$ is the set of $\vec d=(\vec d_a,\vec d_b,\vec d_c)$ which is not invariant under the projection and $\vec d'=(\vec d_a,\vec d_b,\vec d_b)$ the invariant set. One should note that \req{realZpro} is not just a squareroot of \req{Zpro}. 

Let us illustrate \req{realZpro} at hand of the topological vertex geometry. The orientifold is taken to identify $Q_2$ and $Q_3$, while $Q_1$ is mapped to itself, see figure \ref{topvertexgeofig}. From the fundamental relation \cite{Walcher:2007qp}
\beq\eqlabel{nNrelation}
n^{(\h g)}_{\vec d}\equiv N^{(\chi)}_{\vec d_a,\vec d_b}\,\,{\rm mod}\,2\,,
\eq
and \req{GVtopvertex} we immediately deduce that the non-vanishing real Gopakumar-Vafa invariants are
\beq\eqlabel{realGVtopvertex}
\begin{split}
|N^{(-1)}_{0,1}|=|N^{(-1)}_{1,0}|&=1\,,\\
|N^{(-1)}_{1,1}|&=1\,.
\end{split}
\eq
Unfortunately, the exact signs of the invariants are not determined by \req{nNrelation}. Let us utilize the real topological vertex \cite{Krefl:2009md,Krefl:2009mw} to explicitly compute the invariants \req{realGVtopvertex}. The real topological partition function of the topological vertex geometry can be expressed in the vertex formalism as
\beq\eqlabel{Zrealtopvertex}
\t Z^{real}=\sum_{R_1=R_1^t, R_2} (-1)^{(|R_1|-r(R_1))/2+|R_2|} q^{\kappa(R_2)/4} C^{real}_{R_2R_1} \sqrt{C_{R_1^t\emptypar\emptypar} }C_{R_2^t\emptypar\emptypar}  \, Q_1^{|R_1|/2} Q_2^{|R_2|}\,,
\eq
with $\kappa(R)$ as in \req{kappadef}, $C_{R_1R_2R_3}$ the topological vertex (see equation \req{vertexrelation}) and $C^{real}_{R_2R_1}$ the real topological vertex (\cf, \cite{Krefl:2009mw}). In fact, there are several possible consistent sign schemes in \req{Zrealtopvertex}, depending on the use of twisted or non-twisted real vertex and the choice of $\pm 1$ in the $r$-type sign. Accordingly, this translates to different sign schemes of the invariants \req{realGVtopvertex}. However, what one can do is to bring the expression of the partition function \req{realredZpro} (and so \req{realZpro}) to a canonical form by replacing in each factor the $\sigma_{\vec d_a}$ sign by a $\sigma_{\vec d_a,\vec d_b}$ sign and move, in case $N^{(\chi)}_{\vec d_a,\vec d_b}$ has not the same parity as $n^{(\chi+1)}_{\vec d'}$, the sign of the exponent into this sign. For example, in the case of the topological vertex geometry one can arrange that the real topological partition function can be expressed as
\beq\eqlabel{realtopvertexgeoZsigns}
\t Z^{real}=\frac{M(Q_1Q_2,q)M^{real}_{-}(Q_2^2,q)}{M^{real}_{+}(Q_1,q)M(Q_2,q)M^{real}_{-}(Q_1Q_2^2,q)}\,.
\eq
But one should keep in mind that in \req{realtopvertexgeoZsigns} there are more consistent (and also inconsistent) sign choices. The advantage of this canonical form is that it allows a simple heuristic derivation of the qualitative partition function of models with $n^{(g=0)}_{\vec d}=\pm 1$ and $n^{(g>0)}_{\vec d}=0$. Namely, the way \req{realZpro} arises heuristically from \req{Zpro} is that after identification of K\"ahler moduli, one identifies the MacMohan factors which are paired up and the factors which do not pair up are replaced via
\beq\eqlabel{MtorealM}
M(x,q)\rightarrow M^{real}_\sigma(x,q)\,,
\eq
with some sign $\sigma$ .

The upshot is, while we can predict the qualitative form of the product expansion of $\t Z^{real}$ for models with $n^{(g>0)}_{\vec d}=0$ and $n^{(g=0)}_{\vec d}=\pm 1$ via \req{realZpro} (due to relation \req{nNrelation}), we still have to invoke a real topological vertex computation to fix a consistent sign scheme. The examples we are going to discuss have a simple enough sign structure such that we can directly infer $\t Z^{real}$ (and so $\t\Zcal$) in the heuristic fashion described above. Nevertheless, we will discuss them in some more detail in the following sections.

\subsection{Example 1: Resolved conifold}
\label{LVconifold}
The closed topological string partition function of the resolved conifold is given by
\beq
 Z=M(1,q)\, M(Q,q)^{-1}\,,
\eq
with $q=e^{i\lambda}$ and $Q=e^{-t}$, where $t$ denotes the single K\"ahler parameter of the geometry.

\begin{figure}
\begin{center}
\includegraphics[scale=0.3]{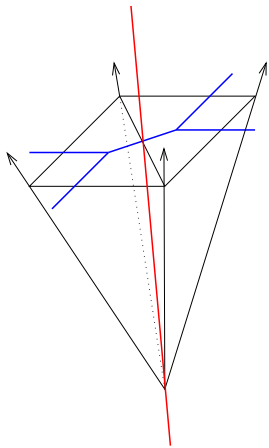}
\end{center}
\caption{Toric fan and web diagram of the resolved conifold with indicated action of the point-reflection (reflection along the red line). }
\label{Conipointref}
\end{figure}

There are two different orientifold projections. Namely, either acting freely or with fixed-points. The action on the toric and web diagram in the fixed-point free case is illustrated in figure \ref{Conipointref}.
In the fixed-point free case, the corresponding partition function has been derived in \cite{Sinha:2000ap} and one can express it in terms of the generalized real MacMohan function as
\beq\eqlabel{ConiZreal}
\t Z^{real}_\pm=M^{real}_\pm(Q,q)^{-1}\,.
\eq 
This trivially confirms the qualitative prediction of the previous section. Note that the origin of the sign degree of freedom can be most clearly seen in the dual Chern-Simons gauge theory on $S^3$ where it corresponds to a choice between $SO$ and $USp$ gauge group \cite{Sinha:2000ap}. 

The case with fixed-points can be dealt with in the real topological vertex formalism of \cite{Krefl:2009md,Krefl:2009mw}. Explicit expansion of the resulting partition function reveals that actually \req{ConiZreal} still holds. This is as expected, since the counting of cross-caps and (odd degree) disks in the conifold should be interchangeable.

With the real Donaldson-Thomas/Gromov-Witten correspondence described in section \ref{realDTinvariants}, we deduce
\beq\eqlabel{ConirealDTZ}
\Zcal_\pm^{real}=M(1,-q)\, M^{real}_{\pm}(Q,-q)^{-1}\,,
\eq 
where we also included the expected degree $0$ contribution $M(1,-q)$ (the action of the involution on the web diagram exchanges the two vertices, see figure \ref{Conipointref}).

\subsection{Example 2: Resolved $\C^2/\Z_n\times\C$}
\label{LVC2Z2C}
\begin{figure}
\psfrag{0}[cc][][0.7]{$0$}
\psfrag{1}[cc][][0.7]{$1$}
\psfrag{2}[cc][][0.7]{$2$}
\psfrag{3}[cc][][0.7]{$3$}
\psfrag{4}[cc][][0.7]{$4$}
\begin{center}
\includegraphics[scale=0.5]{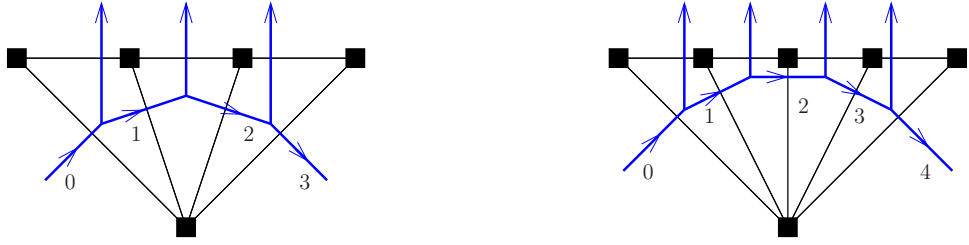}
\end{center}
\caption{Toric diagram and $(p,q)$-web of the resolution of $\C^2/\Z_3\times \C$ (left) and $\C^2/\Z_4\times \C$ (right). The webs are drawn in blue. The numbers correspond to the $k$ label of the edges defined in \req{webparam}.}
\label{Z3andZ4web}
\end{figure}

Let us consider the family of geometries given by the resolution of $\C^2/\Z_n\times \C$ (also known as resolution of local $A_n$). The corresponding toric diagrams can be obtained by subdividing the long edge of the $\C^3$ toric diagram into $n$ pieces of equal length by inserting $n-1$ additional vertices and performing a (unique) triangulation. The corresponding web diagram is the graph dual, and it is easy to see that the edges of the web are given in the $(p,q)$-plane by the set
\beq\eqlabel{webparam}
(p,q)=(1,1-2k/n)\,,
\eq
with $k\in\{0,\dots,n\}$ and $n$ additional edges $(p,q)=(0,1)$. Each edge parameterized by $0<k<n$ correspond to one of the $n-1$ blown-up $\P^1$ in the geometry with K\"ahler modulus $t_{k}$. We define $Q_k:=e^{-t_{k}}$. For illustration, we show the toric and web diagrams for $n=3$ and $n=4$ in figure \ref{Z3andZ4web}.

The closed topological string partition function (without constant map contribution) for these geometries can be easily calculated in the topological vertex formalism of \cite{Aganagic:2003db}. One obtains the expression
\beq
\t Z_n=\sum_{R} q^{-\sum_i \kappa_{R_i}/2} C_{\emptypar \emptypar R_1}\prod_{i=1}^{n-2} C_{R_i^t\emptypar R_{i+1}}\,  C_{R^t_{n-1}\emptypar \emptypar } \,\prod_{i=1}^{n-1} Q^{|R_i|}_i\,,
\eq	
with $R=\{R_1,\dots,R_{n-1}\}$, $\kappa(R)$ as in \req{kappadef} and $C_{R_1R_2R_3}$ the topological vertex (see \req{vertexrelation}).

It can be shown that the partition function $Z_n$ (including the constant map contribution) can be expressed in terms of the generalized MacMohan function $M(x,q)$ as \cite{Young08,Bryan08}
\beq\eqlabel{ZC2ZnxCclosed}
Z_n(Q_k,q)=M(1,q)^{n}\, \prod_{1\leq i\leq j<n} M(Q_{[i, j]},q) \,,
\eq
where we defined 
\beq\eqlabel{Qkomdef}
Q_{[i,j]}=\left\{
\begin{matrix}
Q_iQ_{i+1}\dots Q_j &{\rm for}& i<j\\
Q_iQ_{i-1}\dots Q_j  &{\rm for}& i>j\\
Q_i &{\rm for}& i=j\\
\end{matrix}
\right.\,.
\eq

\begin{figure}
\psfrag{R1}[cc][][0.7]{$R_1$}
\psfrag{R2}[cc][][0.7]{$R_2$}
\psfrag{R3}[cc][][0.7]{$R_3$}
\begin{center}
\includegraphics[scale=0.5]{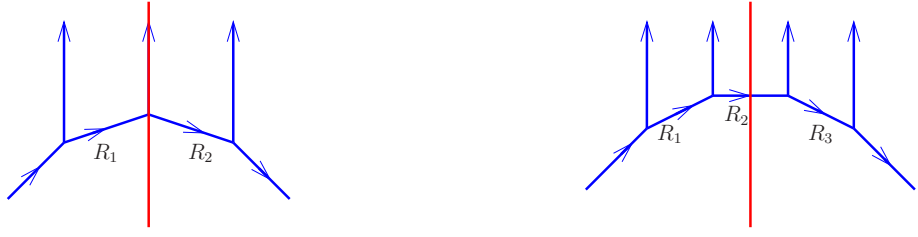}
\end{center}
\caption{$(p,q)$-web of the resolution of $\C^2/\Z_3\times \C$ (left) and $\C^2/\Z_4\times \C$ (right) with representation assoicated to the edges. The action of the involution is indicated via the red line. Note that for $n$ odd we have a single fixed vertex, while for $n$ even a single fixed edge.}
\label{Z3andZ4oweb}
\end{figure}

Let us turn to the real case. The toric, respectively, web diagrams possess a line reflection symmetry which we use to perform an orientifold projection following the formalism developed in \cite{Krefl:2009md,Krefl:2009mw}. In our parameterization \req{webparam} the orientifold simply acts on the geometry via $k\rightarrow n-k$. As is apparent from the action on the web diagrams shown in figure \ref{Z3andZ4oweb}, we have to distinguish between $n$ odd and $n$ even. 
While for $n$ odd there is a single fixed vertex, for $n$ even there is instead a single fixed edge. Thus, for $n$ odd there are no fixed K\"ahler moduli $t_k$ under the projection and $(n-1)/2$ moduli remain in the quotient, while for $n$ even there is a single fixed modulus $t_{n/2}$ and $n/2$ moduli remain. 

We deduce that for $n$ odd the real partition function in the real topological vertex formalism is given by
\beq\eqlabel{Zrealodd}
 \t Z^{real}_{n\, \rm odd} =\sum_R  q^{-\sum_i \kappa_{R_i}/2-\kappa_{R_{(n-1)/2}}/4} C_{\emptypar \emptypar R_1}\prod_{i=1}^{(n-1)/2-1} C_{R_i^t\emptypar R_{i+1}} \, C^{real}_{R^t_{(n-1)/2}\emptypar} \, \prod_{i=1}^{(n-1)/2} Q^{|R_i|}_i\,,
\eq
with $R=\{R_1,\dots,R_{(n-1)/2}\}$, while for $n$ even we have
\beq\eqlabel{Zrealeven}
\t Z^{real}_{n\,{\rm even}}= \sum_R\, q^{-\sum_i^{n/2}\kappa_{R_i}/2}  \, C_{\emptypar \emptypar R_1}\prod_{i=1}^{n/2-1} C_{R_i^t\emptypar R_{i+1}}  \,Q^{|R_{n/2}|/2}_{n/2}\prod_{i=1}^{n/2-1}Q^{|R_i|}_i  \,,
\eq
with $R=\{R_1,\dots,R_{n/2}\}$. Note that for $n$ odd there is the freedom to replace the real vertex by a twisted real vertex while in the $n$ even case we may insert an additional $(-1)^{|R_{n/2}|}$ sign. This sign is similar to the $\pm 1$ freedom of the $r$-type sign of the (freely acting) orientifold of the conifold. 
We will denote the partition functions \req{Zrealodd} and \req{Zrealeven} with real vertex, respectively without extra sign insertion by $Z^{real}_{n,+}$, while with twisted real vertex, respectively with extra sign insertion by $Z^{real}_{n,-}$.

Following the general philosophy of the relation between the closed topological string partition function expressed in terms of generalized MacMohan functions and the corresponding real partition functions discussed in section \ref{simplemodels}, we expect the following product forms to hold
\beq\eqlabel{Zrealoddprod}
\begin{split}
\t Z^{real}_{n\,{\rm odd},\sigma}(Q_k,q)=&\prod_{1\leq i\leq j\leq (n-1)/2}M(Q_{[i,j]},q) \prod_{1\leq i<j\leq (n-1)/2} M(Q_{[i,(n-1)/2]}Q_{[(n-1)/2,j]},q)\\
&\times\prod_{1\leq i\leq (n-1)/2} M^{real}_\sigma(Q_{[i,(n-1)/2]}Q_{[(n-1)/2,i]},q)\,,
\end{split}
\eq
and 
\beq\eqlabel{Zrealevenprod}
\begin{split}
\t Z^{real}_{n\,{\rm even,\sigma}}(Q_k,q)=&\prod_{1\leq i<n/2}M(Q_{[i,n/2]},q)\prod_{1\leq i\leq j<n/2}M(Q_{[i,j]},q)\\
&\times \prod_{1\leq i< j<n/2} M(Q_{[i,n/2]}Q_{[n/2-1,j]},q) \\
&\times M^{real}_\sigma(Q_{n/2},q) \prod_{1\leq i<n/2} M^{real}_\sigma(Q_{[i,n/2-1]}Q_{n/2}Q_{[n/2-1,i]},q)\,.
\end{split}
\eq
We have verified these product forms for small $n$ via low-degree expansion and comparing with the expansions obtained from explicitly evaluating the real vertex expressions \req{Zrealodd} and \req{Zrealeven}.

Note that for $n=2$ (\ie, $\O(-2)\oplus\O(0)\rightarrow\P^1$) we can easily derive $\t Z^{real}_{2,\sigma}$ explicitly. We deduce from \req{Zrealeven} (recall that $C_{R\emptypar\emptypar}=q^{\kappa_R/2}s_{R^t}(q^\rho)$, with $\rho=(-\frac 1 2,-\frac 3 2,\dots)$)
\beq
\t Z^{real}_{2,\sigma}=\sum_\lambda  (-1)^{\frac{(1-\sigma)\lambda|}{2}}Q^{|\lambda|/2}s_{\lambda^t}(q^\rho)=\sum_\lambda (-1)^{\frac{(1-\sigma)|\lambda|}{2}} s_{\lambda}(-Q^{1/2} q^{-\rho})\,.
\eq
Invoking the Schur function identity \req{SchurIDo1} we arrive at
\beq
\t Z^{real}_{2,\pm}=\prod_{i=1}^\infty\frac{1}{1\pm Q^{1/2}q^{i}}\prod_{1\leq i<j}^\infty\frac{1}{1-Qq^{i+j-1}}=M_\pm^{real}(Q,q)\,,
\eq
which translates to the real Donaldson-Thomas partition function
\beq\eqlabel{ZC2Z2final}
\t \Zcal^{real}_{2,\pm}=M(1,-q)\,M_\pm^{real}(Q,-q)\,,
\eq
where we already included the degree $0$ contribution (see below).

The reduced real partition functions $\t Z'$, as defined in \req{reducedZtop}, are given by
\beq\eqlabel{orbredrealZ1}
\t Z'_{n\,{\rm odd},\sigma}=\prod_{1\leq i\leq (n-1)/2} M'_\sigma(Q_{[i,(n-1)/2]}Q_{[(n-1)/2,i]},q)\,,
\eq
and
\beq\eqlabel{orbredrealZ2}
\t Z'_{n\,{\rm even},\sigma}=M'_\sigma(Q_{n/2},q) \prod_{1\leq i<n/2} M'_\sigma(Q_{[i,n/2-1]}Q_{n/2}Q_{[n/2-1,i]},q)\,.
\eq
The corresponding real Donaldson-Thomas partition functions can be obtained by simply substituting $q\rightarrow -q$ in the above expressions. Finally, note that since $|V^a|\equiv n\,{\rm mod}\, 2$ and $|V^b|/2=\lfloor n/2\rfloor$, as can be easily inferred from figure \ref{Z3andZ4oweb}, and following the degree $0$ discussion of section \ref{d0contribution}, we expect that
\beq\eqlabel{constmap}
{\Zcal^0}^{real}_{n,\sigma}= M^{real}_\sigma(1,-q)^{n\,{\rm mod}\,2}  M(1,-q)^{\lfloor n/2\rfloor}\,,
\eq
captures the degree $0$ contribution in the real case.

\section{Orbifold point}
\label{orbifoldpoint}
In this section we are going to discuss a real version of orbifold Donaldson-Thomas invariants, whose definition is particularly simple. The main class of examples we are going to discuss are the orbifolds $\C^2/\Z_n\times\C$, for which we will derive the real orbifold partition functions in a combinatorial fashion and compare with the large volume partition functions derived in the previous section.
 
\subsection{Real orbifold Donaldson-Thomas invariants}
\label{orbidef}
One can define Donaldson-Thomas invariants of orbifolds $\C^3/G$ with $G$ a finite subgroup of $SU(3)$ as follows \cite{Young08} (we restrict to the case with $G$ abelian). We define 
\beq
\Hilb^R(\C^3/G)=\{Y\subset \C^3:\,{\rm Y\, is}\, G{\rm-invariant},\, H^0(\O_Y)=R \}\subset \Hilb^n(\C^3)\,,
\eq
with $R$ a $n$-dimensional representation of $G$. The $\T$ action of $\C^3$ commutes with $G$ and therefore induces a $\T$ action on $\C^3/G$ and $\Hilb^R(\C^3/G)$. One can use Behrend's $\nu$-function \cite{Behrend05,BF05} to define the orbifold Donaldson-Thomas invariant as a weighted euler characteristic of the Hilbert scheme, \ie,  
\beq\eqlabel{oinv}
\c d_{R}=\chi(\Hilb^R(\C^3/G),\nu)=\sum_{\alpha}(-1)^{\dim T_\alpha \Mcal}\,,
\eq
where $T_\alpha \Mcal$ denotes the Zariski tangent space of $\Hilb^R(\C^3/G)$ at the fixed-point $\alpha\in \Hilb^R(\C^3/G)$ under the $\T$ action. Similar as in \req{monideal}, each fixed-point $\alpha$ can be identified with a 3d partition, which however one can now see as a representation of both, the $\T$ and $G$ action. Let us denote the irreducible representations of $G$ (which are 1 dimensional) by $R_i$. Hence, $R=\sum_i  d_i R_i$. To each irreducible representation we associate a variable $q_i$ and with the abbreviation $q^R=q^{d_0}_0\dots q^{d_r}_r$ we define the orbifold Donaldson-Thomas partition function as
 \beq\eqlabel{orbiDTZ}
 \c \Zcal=\sum_R \c d_R\, q^R\,.
 \eq
At least in certain cases, the partition function $\c\Zcal$ can be explicitly evaluated combinatorially \cite{Young08}. We will come back to this at hand of $G=\Z_n$ in section \ref{orbifoldsec}. 

The above definition of orbifold Donaldson-Thomas invariants changes in the real case as follows. We define
\beq\eqlabel{oHilb}
\begin{split}
\Hilb_\inv^{R}(\C^3/G)&:=\{Y\subset \C^3:\,{\rm Y\, is}\, G\,{\rm and}\, \inv {\rm-invariant},\, H^0(\O_Y)=R \}\\
&\subset \Hilb^{R}(\C^3/G)\subset \Hilb^{n}(\C^3)\,,
\end{split}
\eq
with $\inv$ as in \req{invdef}. The real orbifold Donaldson-Thomas invariants are defined to be
\beq\eqlabel{realoinv}
\c D_{R}=\sum_{\alpha}(-1)^{\frac{\dim T_\alpha \Mcal}{2}}\sigma_\alpha\,,
\eq
where $\alpha$ runs over the fixed points of $\T'\subset \T$ acting on $\Hilb_\inv^R(\C^3/G)$ and $\sigma_\alpha$ a model dependent sign (which is however trivial for the class of models we will discuss in section \ref{orbifoldsec}).  Let us split the irreducible representations of $G$ into two sets, $R_{a_i}$ and $R_{b_i}$, where the  $R_{a_i}$ are identified with themselves and the $R_{b_i}$ are identified pairwise under $\inv$ (we take the $R_{b_i}$ to be in the quotient in the following). Then, $R=\sum_i d_{a_i}R_{a_i}+2\sum_j  d_{b_j}R_{b_j}$. Associating the variables $\pm q^{1/2}_{a_i}$ with $R_{a_i}$ and $\pm q^{1/2}_{b_i}$ with $R_{b_i}$, we define the real orbifold partition function as
\beq\eqlabel{realorbZ}
\c \Zcal^{real}_\pm=\sum_{R_a, R_b}\, (\pm 1)^{R_a}\, \c D_{R_a,R_b}\,q^{R_a/2} q^{R_b}\,.
\eq
A hint that the above made definitions are indeed what we are after comes from the fact that we obtain for $G=1$ and $\sigma_\alpha=1$,
\beq
\c\Zcal^{real}_\pm=\Zcal_\pm^{real}\,,
\eq
as it should be. Finally, note that the total ``charge" $R$ may now be fractional, \ie, it can take integer and half-integer values. 

\subsection{Review of transfer matrix approach}
\label{transfermatrix}
In the explicit examples to be discussed in the following sections, we have to evaluate the partition functions of certain combinatorial systems. At least in specific cases, this can be conveniently achieved via the transfer matrix approach of \cite{Okounkov:2003sp,OR05}, which we briefly review here for completeness.

The transfer matrix approach is best suited for combinatorial setups which can be viewed as a stack of 2d partitions (integer partitions), where each element of a 2d partition carries the same weight $q_g$, however not necessarily constant for all 2d partitions. A 2d partition is usually represented by a Young (also known as Ferrers) diagram, and a stack is a sequence of Young diagrams $\lambda^{(i)}$, which interlace in a specific manner. Two 2d partitions are said to interlace, denoted as $\lambda \succ \mu$, if they satisfy
\beq
\lambda_1\geq\mu_1\geq\lambda_2\geq\mu_2\geq\dots\,,
\eq
with $\lambda_i$ and $\mu_i$ the $i$th column of the partition.  One may view the different weights $q_g$ associated to the Young diagrams as a coloring of the stack. 

The partition function of the setup can be evaluated as follows. We associate to each 2d partition $\lambda$ a state $\ket{\lambda}$ in the Hilbert space of a complex fermion. Via bosonization, we introduce the operators $\Gamma_\pm(q)$ and $\Gamma_\pm'(q)$ given by \cite{Kac}
\beq\eqlabel{Gopdef}
\begin{split}
\Gamma_\pm(q)&=\exp{\sum_{k>0} \frac{q^k}{k}\alpha_{\pm k}}\,,\\
\Gamma'_\pm(q)&=\exp{\sum_{k>0} \frac{(-1)^{k-1}q^k}{k}\alpha_{\pm k}}\,,\\
\end{split}
\eq
where $\alpha_{\pm k}$ are bosonic annihilation and creation operators satisfying the commutation relation $[\alpha_n,\alpha_{-m}]=n\delta_{n,m}$. As can be easily inferred from the definition \req{Gopdef}, the $\Gamma$ operators fulfill the following commutation relations (see for instance \cite{Young08})
\beq\eqlabel{Gcommute}
\begin{split}
\Gap{x}\Gam{y}&=\left(\frac{1}{1-xy}\right)\Gam{y}\Gap{x}\,,\,\,\,\,\,\,\,\,\Gap{x}\Gamp{y}=\left(1+xy\right)\Gamp{y}\Gap{x}\,,\\
\Gapp{x}\Gamp{y}&=\left(\frac{1}{1-xy}\right)\Gamp{y}\Gapp{x}\,,\,\,\,\,\,\,\,\, \Gapp{x}\Gam{y}=\left(1+xy\right)\Gam{y}\Gapp{x}\,.\\
\end{split}
\eq
Especially, applied to a state of the Hilbert space $\ket{\lambda}$, these operators yield for $q=1$ 
\beq\eqlabel{G1ops}
\begin{split}
\Gam{1}\ket{\lambda}=\sum_{\mu\succ\lambda}\ket{\mu} &\,, \,\,\,\,\,\,\,\,\,\, \Gap{1}\ket{\lambda}=\sum_{\mu\prec\lambda}\ket{\mu}\,, \\
\Gamp{1}\ket{\lambda}=\sum_{\mu^t\succ\lambda^t}\ket{\mu} &\,, \,\,\,\,\,\,\,\,\,\, \Gapp{1}\ket{\lambda}=\sum_{\mu^t\prec\lambda^t}\ket{\mu}\,. \\
\end{split}
\eq
That is, they generate all states $\ket{\mu}$ which correspond to a 2d partition $\mu$ interlacing with $\lambda$. Furthermore, we introduce operators $\h q_g$ with the property that $\h q_g\ket{\lambda}=q_g^{|\lambda|}\ket{\lambda}$, where $|\lambda|$ denotes the number of elements (or boxes) of the 2d partition $\lambda$. Note that the operators $\h q_g$ commute with the $\Gamma$ operators as follows \cite{Young08}
\beq\eqlabel{qcommute}
\begin{split}
\Gap{x}q_g=q_g\Gap{x q_g}\,, &\,\,\,\,\,\,\,\,\,\, q_g\Gam{x}=\Gam{x q_g}q_g\,,\\
\Gapp{x}q_g=q_g\Gapp{x q_g}\,, &\,\,\,\,\,\,\,\,\,\, q_g\Gamp{x}=\Gamp{x q_g}q_g\,.
\end{split}
\eq
We conclude from the relations \req{G1ops} that via applying the operators $\Gamma(1)$ and $\h q_g$ in an appropriate order to the vacuum (represented by the state $\ket{\emptypar}$), we obtain the partition function of the corresponding combinatorial system. Hence, in order to obtain the explicit partition function, we just have to evaluate a correlator
\beq\eqlabel{basiccorrelator}
\bracket{\emptypar}{\dots}{\emptypar}\,,
\eq
where the dots stand for the insertion of (possibly infinitely many) $\Gamma$ and $\h q_g$ operators with order determined by the interlacing pattern. Let us assume that there is at least one local maximum (or minimum) in the stack of Young diagrams, that is, there is a $\lambda^{(i)}$ with weight $q_i$ in the stack for which $\lambda^{(i-1)}\prec \lambda^{(i)}\succ \lambda^{(i+1)}$, respectively, $\lambda^{(i-1)}\succ \lambda^{(i)}\prec \lambda^{(i+1)}$ holds. With $\h q_i$ the corresponding weight operator, the correlator \req{basiccorrelator} takes the form
\beq
\bracket{\emptypar}{\dots \h q_i\dots}{\emptypar}\,.
\eq
It is convenient to define a left and right state, $\bra{\Omega_L}$, respectively, $\ket{\Omega_R}$, via
\beq\eqlabel{corrOmega}
\bracket{\Omega_L}{\h q_i^{-1/2}\h q_i^{1/2}}{\Omega_R}:=\bracket{\emptypar}{\dots \h q_i\dots}{\emptypar}\,.
\eq
A maximum can be ``blown up" by inserting additional partitions $\lambda^{(i,j)}$ with $j\in\{1,\dots,N-1\}$ with interlacing pattern
\beq\eqlabel{blowupinterlace}
\lambda^{(i-1)}\prec\lambda^{(i,0)}\succ\lambda^{(i,1)}\prec\lambda^{(i,2)}\succ\dots\prec\lambda^{(i,N-2)}\succ\lambda^{(i,N-1)}\prec\lambda^{(i+1)}\,,
\eq
where we defined $\lambda^{(i,0)}:=\lambda^{(i)}$. A similar blow up can be performed for a minimum. We denote the operator which implements the blow up in the correlator as $\h \Wcal$ and hence \req{corrOmega} reads after the blow up
\beq
\bracket{\Omega_L}{\h\Wcal^{N-1}}{\Omega_R}\,.
\eq
For reasons that will become clear later, $\h\Wcal$ is referred to as wall crossing operator \cite{Sulkowski:2009rw}.

 As we will see more explicitly in the examples below, in the evaluation of the correlators the following identities relating the $\Gamma$ operators to skew Schur functions are particular useful
\beq\eqlabel{statetoSchur}
\begin{split}
\prod_{i=1}\Gam{x_i}\ket{\lambda}=\sum_\mu s_{\mu/\lambda}(x)\ket{\mu}\,, &\,\,\,\,\,\,\,\,\,\,  \bra{\lambda}\prod_{i=1}\Gap{x_i}=\sum_\mu \bra{\mu} s_{\mu/\lambda}(x)\,,    \\
\prod_{i=1}\Gamp{x_i}\ket{\lambda}=\sum_\mu s_{\mu^t/\lambda^t}(x)\ket{\mu}\,, &\,\,\,\,\,\,\,\,\,\, \bra{\lambda}\prod_{i=1}\Gapp{x_i}=\sum_\mu \bra{\mu} s_{\mu^t/\lambda^t}(x)\,.  \\
\end{split}
\eq

If the combinatorial arrangement possesses a $\Z_2$ symmetry which is compatible with the representation of the setup as a stack of Young diagrams, we can use the transfer matrix formalism as well to calculate the partition function of the $\Z_2$ invariant subset \cite{Krefl:2009mw}. Especially, the symmetry requires that a local maximum or minimum exists which is fixed under the $\Z_2$. In the examples we are going to discuss below, we have to distinguish between two cases. Either the symmetry leaves the central Young diagram in the stack invariant, or it acts on it as a reflection along the diagonal (\cf, figure \ref{2dpartitions}). In the latter case the central Young diagram needs to be self-conjugate, \ie, $\lambda=\lambda^t$. It follows that in order to obtain the quotient partition function, in the first case we simply have to sum over all possible central Young diagrams, that is, the projected correlator is given by 
\beq\eqlabel{Plineref}
\sum_\lambda (-1)^{\frac{(1-\sigma)|\lambda|}{2}}\bracket{\lambda}{\h q_i^{1/2}}{\Omega_R}\,,
\eq
while for the second case we have
\beq\eqlabel{Ppointref}
\sum_{\lambda=\lambda^t}(-1)^{\frac{(1-\sigma)|\lambda|\pm r(\lambda)}{2}} \bracket{\lambda}{\h q_i^{1/2}}{\Omega_R}\,,
\eq
where $r(\lambda)$ denotes the number of diagonal boxes of the Young diagram $\lambda$. In both cases we allowed for an additional sign weighting $(-1)^{|\lambda|}$ (parameterized by $\sigma$) of the central 2d partition. Note the insertion of the extra sign $(-1)^{\pm r(\lambda)/2}$ in \req{Ppointref}. The reason for this sign insertion will be explained in section \req{conisec}.

Clearly, we can still blow up the maximum, respectively minimum in the projected correlator. For that, we need to split the operator $\h\Wcal^{N-1}$ into a left and right part, \ie,
\beq\eqlabel{Wsplit}
\h\Wcal^{N-1}=\h\Wcal_L\, \h\Wcal_R\,,
\eq
and only $\h\Wcal_R$ enters \req{Plineref} and \req{Ppointref}. Note that the splitting \req{Wsplit} is different for $N-1$ even and odd, as is apparent from \req{blowupinterlace}, and will become more clear in the explicit examples to be discussed in section \req{conisec} and \req{c2z2Nsec}.

\subsection{Example: $\C^2/ \Z_n\times \C$ orbifold}
\label{orbifoldsec}
Let us now consider our main example for the orbifold point, namely the $\C^2/\Z_n\times\C$ orbifold. The orbifold Donaldson-Thomas partition function for this model has been derived in \cite{Young08} via translating the calculation to a combinatorial problem. The essential ingredients for this translation are that one can identify each $\T$ fixed point with a $\Z_n$-colored plane partition and that $\dim T_\alpha \Mcal=|\pi|_0$ \cite{Young08}. Thus, from the definitions \req{oinv} and \req{orbiDTZ} one infers
\beq\eqlabel{colcryspart}
\c \Zcal_n(-q_0,q_1,\dots)\equiv P_n(q_k)=\sum_{\pi\in\Pcal_n} \prod_{k=0}^{n-1} q_{k}^{|\pi|_k}\,,
\eq
where $\Pcal_n$ is the set of $\Z_n$-colored plane partitions, $q_k$ are weights and $|\pi|_k$ denotes the number of boxes of color $k$  in $\pi$. Recall that a $\Z_n$-colored 3d partition is a 3d partition with boxes colored with $n$ colors such that addition in $\Z^3_{\geq 0}$ respects the group law of $\Z_n$. Note that a choice of coloring of the unit vectors (and null-vector) uniquely determines a coloring of the partition. We choose that the boxes at $(0,0,0)$ and $(0,0,1)$ carry color $0$ (the identity element of $\Z_n$) and the boxes at $(1,0,0)$ and $(0,1,0)$ color $1$, respectively $n-1$ (the inverse of $1$). In this coloring scheme, the partition is monochromatic in the $z$-direction, while in the $(x,y)$-plane, it is monochromatic with color $k$ along the rays $y-x=k$. This means that in diagonal slicing the $\Z_n$-colored 3d partition decomposes into monochromatic slices and we can apply the transfer matrix approach to compute the partition function as outlined in section \ref{transfermatrix}.

As shown in \cite{Young08}, the orbifold partition function can be obtained from the large volume partition function (given in equation \req{ZC2ZnxCclosed} up to $q\rightarrow -q$), via the substitution 
\beq\eqlabel{closedsub}
M(x,-q)\rightarrow\widetilde{M}(x,-q)=M(x,-q)\, M(x^{-1},-q)\,,
\eq 
and performing the reparameterization 
\beq\eqlabel{orbireparam}
Q_k\rightarrow q_k,\,\,\,\,\, q\rightarrow q_{[0,n-1]}\,.
\eq
That is, as already outlined in the introduction, the partition function at the large volume point is connected to the reparameterized partition function at the orbifold point via a wall crossing factor as in \req{Wfactor}, \ie,
\beq
 \c\Zcal_n(Q_k,q)=\Wcal \,\Zcal_n(Q_k,q) = \Zcal_n(Q^{-1}_k,q)\, \t \Zcal_n(Q_k,q)\,,
\eq
where $ \c\Zcal_n(Q_k\rightarrow q_k,q\rightarrow q_{[0,n-1]})\equiv\c \Zcal_n(q_k)$. One should note that the reparameterized partition function can be given a BPS state counting interpretation in terms of D6-D0 bound states, \ie, $\c\Zcal_n(Q_k,q)\equiv Z_{BPS}^{(1)}$, where however the D0-branes are fractional branes, that is, they can carry D0 and D2 charge. The origin of the D2 charge are D2- branes wrapped on 2-cycles which become point-like at the orbifold point.

\begin{figure}
\psfrag{x}[cc][][1]{$x$}
\psfrag{y}[cc][][1]{$z$}
\psfrag{z}[cc][][1]{$y$}
\begin{center}
\includegraphics[scale=0.3]{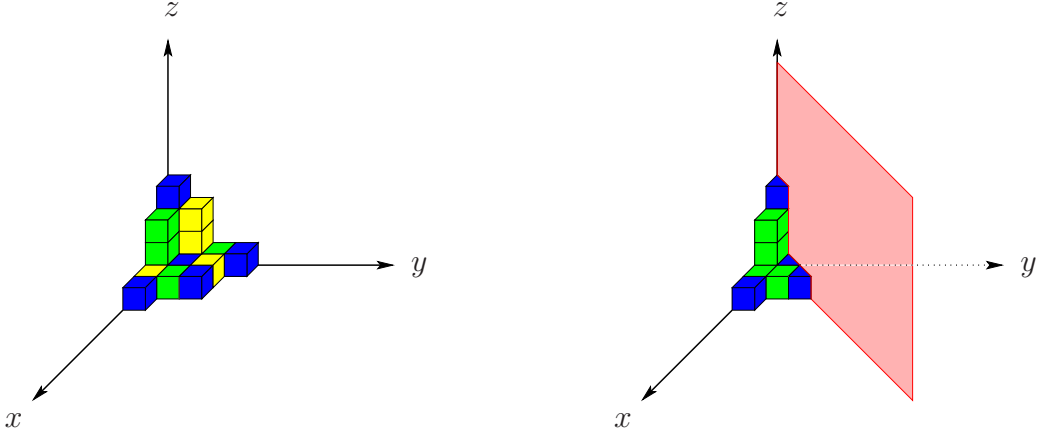}
\end{center}
\caption{Left: $3$-colored symmetric plane partition. Right:  After identification of (two) colors and taking the quotient.}
\label{C2Z3crystal}
\end{figure}

Let us now consider the additional action of the orientifold. Since the $\Z_n$ group action on $\C^2$ is given by
\beq
(x,y)\rightarrow (\omega^i x,\omega^{-i} y)\,,
\eq
with $\omega=e^{2\pi\ii /n}$ and $i\in\{0,\dots,n-1\}$, hence commutes with the orientifold action 
\beq
\inv: (x,y)\rightarrow (\cc y,\cc x)\,,
\eq
we can obtain the orbifold partition function in the real case in a similar fashion as above. Namely, via calculating the partition function of a $\Z_n$-colored symmetric 3d partition (each symmetric partition corresponds to a fixed-point surviving the orientifold projection) and identifying colors according to the $\inv$ action, that is, setting $q_k=q_{k-n}$. Thus, 
\beq\eqlabel{realcolcryspart}
\c\Zcal_{n,\sigma}^{real}(-q_0,q_1,\dots)\equiv R_{n,\sigma}(q_k)=\sum_{\mathcal \pi\in \mathcal S_n} (-1)^{\frac{(1-\sigma) \tr(\pi)}{2}}\prod_{k=0}^{n-1}q_k^{|\pi|_k/2}\,,
\eq
where $\mathcal S_n\subset \mathcal P_n$ denotes the set of symmetric $\Z_n$-colored 3d partitions, up to identification of colors.  For illustration, a colored symmetric 3d partition (after identification of colors) entering the real orbifold partition function of $\C^2/\Z_3\times \C$ is shown in figure \ref{C2Z3crystal}.
We inserted an additional factor of $1/2$ in the exponent because we take the quotient and an additional sign weighting of the diagonal 2d partition to account for the sign freedom as in \req{realorbZ}. However, we inserted no extra sign to account for $\sigma_\alpha$ in \req{realoinv}, \ie, we set $\sigma_\alpha=1$. As we will see below, this sign choice yields the expected result. 

The partition function \req{realcolcryspart} can be explicitly calculated via the transfer-matrix approach outlined in section \ref{transfermatrix}. In detail, we have to evaluate the correlator
\beq\eqlabel{C2Znbasiccorr}
R_{n,\sigma}=\sum_\lambda(-1)^{\frac{(1-\sigma )|\lambda|}{2}}\bracket{\lambda}{\h q_0^{1/2}\prod^\infty\left(\Gam{1} \h q_{n-1}\Gam{1}\h q_{n-2}\dots\Gam{1}\h q_{0}\right)}{\emptypar}\,,
\eq
and identifying colors afterwards. 
Commuting the $q_g$ operators to the right using the relations \req{qcommute} we obtain
\beq\eqlabel{C2ZnCcor}
\begin{split}
R_{n,\sigma}&=\sum_\lambda(-1)^{\frac{(1-\sigma) |\lambda|}{2}}\\
&\times\bracket{\lambda}{\prod^\infty_{i=1}\left(\Gam{q_1^{-1/2}\dots q^{-1/2}_{n-1} q^{i-1/2}}\dots\Gam{q_1^{1/2}\dots q_{n-1}^{1/2}q^{i-1/2}}\right)}{\emptypar}\,,
\end{split}
\eq
where we set $q=q_0q_1\dots q_{n-1}$. We can express this correlator entirely in terms of Schur functions with help of the relations \req{statetoSchur} and then use Schur function identities to solve it. The calculation is straight-forward, however rather lengthy and cumbersome. Therefore we give the details of the calculation in appendix \ref{C2ZNxCcalc} and here just quote the result. That is, from the calculation in appendix \ref{C2ZNxCcalc} we deduce that the reparameterized real Donaldson-Thomas partition function at the orbifold point can be obtained via substituting \req{closedsub} and 
\beq\eqlabel{realsub}
M^{real}_\pm(x,-q)\rightarrow\widetilde{M}^{real}_\pm(x,-q)=M^{real}_\pm(x,-q) M^{real}_\pm(x^{-1},-q)\,,
\eq
in the large volume partition functions \req{Zrealoddprod} and \req{Zrealevenprod}. This fits nicely with the general expectation \req{realWC}, \ie, the real partition functions are related via a wall crossing factor of $\Wcal=\t \Zcal_n^{real}(Q_k^{-1},q)$. The corresponding reduced wall crossing factor $\Wcal'=\t \Zcal_n'(Q_k^{-1},q)$ can be easily deduced from \req{orbredrealZ1} and \req{orbredrealZ2}.

Finally, note that the combinatorial derivation allows us to identify the degree 0 contribution. We deduce (see appendix \req{C2ZNxCcalc}) that the degree $0$ contribution is as stated in equation \req{constmap}.

\section{Non-commutative point}
\label{NCpoint}
In this section we will define real non-commutative Donaldson-Thomas invariants. As examples, we will discuss the non-commutative resolutions of the conifold and of $\C^2/\Z_2\times\C$. Especially, we will derive the corresponding real non-commutative Donaldson-Thomas partition functions and compare to the real large volume partition functions.

\subsection{Real non-commutative Donaldson-Thomas invariants}
\label{defNCDT}

Consider a quiver (connected oriented graph) $Q=\{V,E\}$ with superpotential $W$.  The set $V$ consists of vertices $v_i\in V$ and oriented edges $x_{ij}\in E$ connecting vertices $v_i$ and $v_j$. The set of oriented paths formed by the edges defines a $\C$-algebra, with product given by concatenation of oriented paths. We refer to the algebra as the path algebra $\C Q$. The superpotential is a linear combination of closed oriented paths (loops) an therefore is an element of the path algebra, \ie, $W\in \C Q$. We can define a differentation $\d_{x_i}W$ by taking each term in $W$ containing $x_i$, rotating it cyclically until $x_i$ is at the first position and then delete $x_i$. We require that each edge $x_i$ occurs exactly once in two different terms of $W$ corresponding to loops with opposite orientation (the orientation of a loop is reflected in $W$ by a $\pm 1$ sign of the corresponding term). This condition ensures that we can construct a brane tiling (a $\T^2$ periodic infinite bipartite graph) that encodes the quiver $Q$ with superpotential $W$ (\cf, appendix A of \cite{Franco:2007ii}). The vertices of the quiver map to faces, edges to edges and the superpotential terms to loops encircling the vertices of the brane tiling (for a detailed introduction to brane tilings we refer to \cite{Kennaway:2007tq}). Since the superpotential defines an ideal $I_W=\left<\d W\right>$, we can define the quotient algebra
\beq
A=\C Q/I_W\,.
\eq
Note that we will denote the quiver with relations as $\Gamma:=(Q,I_W)$. Under certain conditions (for details we refer to the mathematics literature), $A$ is a smooth non-commutative 3-Calabi-Yau algebra and the spectrum of the center of the algebra $Z(A)$ is a toric Calabi-Yau 3-fold singularity,
\beq
X=\spec \, Z(A)\,.
\eq
Especially, the algebra $A$ is a non-commutative crepant resolution of $X$ \cite{Bergh}. This resolution is what we mean by non-commutative point in K\"ahler moduli space.

A representation of $\Gamma$ is specified by associating to each $v_j$ a vector space $U_j$ of dimension $n_j$ and to each edge $x_{ij}$ a linear map $U_i\rightarrow U_j$ satisfying the relations $I_W$. We encode the dimensions of the vector spaces in the dimension vector $\vec n\in\N^{|V|}$. In order to be able to construct a moduli space of representations of a quiver, we need to introduce a notion of $\theta$-stability \cite{King}. Therefore, let us associate to the vertices real numbers $\theta_j$ which satisfy the condition $\vec \theta \cdot\vec n=0$ for a given dimension vector $\vec n$, where $\vec \theta:=(\theta_1,\theta_2,\dots)$. A representation of $\Gamma$ is called semi-stable, if for every proper subrepresentation with dimension vector $\vec n'$, we have that $\vec \theta\cdot \vec n'\leq 0$, and stable if $\vec\theta\cdot\vec n'<0$. Let us choose one vertex $v_i$ and add a new vertex $v_*$ connected with a single edge $x_{*i}$ to $v_i$. This yields a new dimension vector which we take to be given by ${(n_*=1,\vec n)}\in \N^{|Q|}\times\N$ and a new stability $\theta'$ taken to be $(\theta_*>0,\vec \theta)\in  \R^{|Q|}\times\R$. Since we made the choice $n_*=1$ (physically, corresponding to introducing a single $D6$ brane) and $\theta_*>0$, stability is equivalent to cyclicity, in the sense that a vector in $U_*$  (or equivalently, a vector in $U_i$) generates the entire representation. Thus, we have that the moduli space of cyclic representations of the quiver $Q$ with relations $\left<\d W\right>$ is given by
\beq
\Mcal_{\vec n,i}=\frac{\left\{
x_{kl}\in \Hom(U_k,U_l), u\in U_i: x_{kl}\,{\rm satisfy}\, \left<\d W\right>, u\,{\rm generates}\, \oplus_j U_j
\right\}}{\prod_j {\rm GL}(U_j)}\,.
\eq
Following \cite{Szendroi07,Mozgovoy:2008fd}, we define a notion of non-commutative Donaldson-Thomas invariants as follows. We define similar as in the orbifold case the non-commutative Donaldson-Thomas invariant via Behrend's $\nu$-function as the weighted euler characteristic
\beq
\c d_{i,\vec n}=\chi(\Mcal_{\vec n,i},\nu)=\sum_{\alpha}(-1)^{\dim T_\alpha\Mcal}\,,
\eq
where the last equality follows from localization with respect to a torus action on $\Mcal_{\vec n,i}$ inherited from the $\T$ action of the toric background. The sum runs over the $\T$ fixed points and $T_\alpha\Mcal$ is the Zariski tangent space of $\Mcal_{\vec n,i}$ at the fixed-point $\alpha$. We assemble the invariants $\c d_{i,\vec n}$ into the non-commutative Donaldson-Thomas partition function
\beq\eqlabel{NCDTZ}
\c \Zcal_i=\sum_{\vec n} \c d_{i,\vec n}\, q^{\vec n}\,,
\eq
with $q^{\vec n}=q_1^{n_1}q_2^{n_2}\dots q_{|V|}^{n_{|V|}}$, where $q_i$ is a variable associated to the $i$th vertex. Note that the invariants and so the partition function explicitly depend on the choice of vertex $v_i$. Since we will only consider fully symmetric examples, we will drop this index in the following sections.

As discussed in detail in \cite{Mozgovoy:2008fd}, each fixed-point $\alpha$ corresponds to a perfect matching $\Delta_\alpha$ of the brane tiling corresponding to $\Gamma$ (recall that a perfect matching is a collection of edges of a brane tiling with the property that this set of edges cover all vertices in a way that each vertex is covered only by a single edge), which asymptotically approaches the empty room matching $\Delta^0$. (The empty room matching is the matching which alignes with the web diagram). Hence, we can explicitly calculate the invariants $\c d_{i,\vec n}$ by enumerating perfect matchings. As shown in \cite{Ooguri:2008yb}, there is a one-to-one correspondence between a perfect matching and a configuration of a crystal. Hence, one can as well compute \req{NCDTZ} by solving a combinatorial problem, which was also the original approach of \cite{Szendroi07} for the non-commutative conifold. 

\begin{figure}
\begin{center}
\psfrag{1}[cc][][0.7]{$x_1$}
\psfrag{2}[cc][][0.7]{$x_2$}
\psfrag{3}[cc][][0.7]{$x_3$}
\includegraphics[scale=0.3]{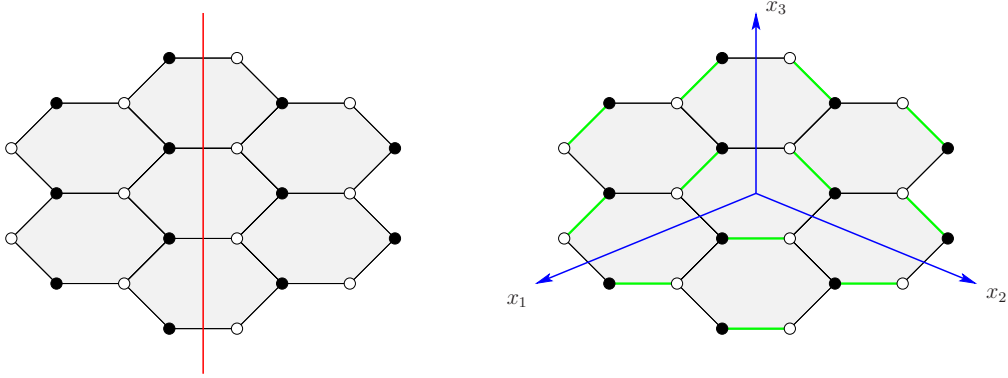}
\end{center}
\caption{Left: Brane tiling corresponding to the $\C^3$ quiver. One of the $\Z_2$ reflection symmetries is indicated via the red line. Right: The perfect matching corresponding to the empty room configuration. The web diagram is indicated via the blue lines. Note that each edge of the web diagram correspond to a path $x_i^n\in\C Q$ (to a mesonic operator in physical language).}
\label{C3dimer}
\end{figure}

In the real case, we define real non-commutative Donaldson-Thomas invariants similar as for the large volume and orbifold point via localization with respect to the $\T'$ invariant under the $\inv$ action. Hence,
\beq\eqlabel{realNCDTinv}
\c D_{i,\vec n_a,\vec n_b}=\sum_{\alpha}(-1)^{\frac{\dim T_\alpha\Mcal}{2}} \sigma_\alpha\,,
\eq
where the sum now runs over the fixed-points $\alpha$ of $\T'$ and we parameterized an additional sign weighting of each fixed-point contribution by $\sigma_\alpha$. Further, we split the dimension vector $\vec n$ into a part $\vec n_a$ invariant under the action of the orientifold and a non-invariant part $\vec n_b$, where $\vec n_b$ is taken in the quotient. In quiver language, the reason for this splitting is that the projection on the quiver identifies some of the vector spaces $U_j$ with themselves while others are identifed with each other. Thus, we have the partition function
\beq
\c \Zcal_{i,\sigma}^{real}=\sum_{\vec n_a,\vec n_b} (\pm 1)^{\vec n_a} \c D_{i,\vec n_a,\vec n_b}\, q^{\vec n_a/2} q^{\vec n_b}\,.
\eq 
Since the fixed-points of the $\T$ action are in one-to-one correspondence to perfect matchings of a brane tilling, it is clear that the fixed-points of the $\T'$ action are in correspondence to perfect matchings which are symmetric with respect to the action of the orientifold projection on the brane tilling.  It thus remains to deduce the action of $\inv$  on the brane tilling (from which one can then easily infer as well the action on the corresponding crystal). But this is simple. We know how $\inv$ acts on the coordinates of the geometry. Combined with the fact that the coordinates are encoded in the quiver as mesonic operators, which in turn are specific paths in the brane tilling, we can directly infer the action of $\inv$ on the tilling (see for instance the extensive discussion in \cite{Franco:2007ii}). For example, we illustrate the pure $\C^3$ case in figure \ref{C3dimer}.
It is straight-forward to translate the action on the tilling to an action on the corresponding crystal configuration. In the remaining sections we will discuss two explicit examples. Especially, for the conifold we will observe that there is a non-trivial sign $\sigma_\alpha$ in \req{realNCDTinv}.

\subsection{Example 1: Non-commutative conifold}
\label{conisec}
\begin{figure}
\begin{center}
\includegraphics[scale=0.5]{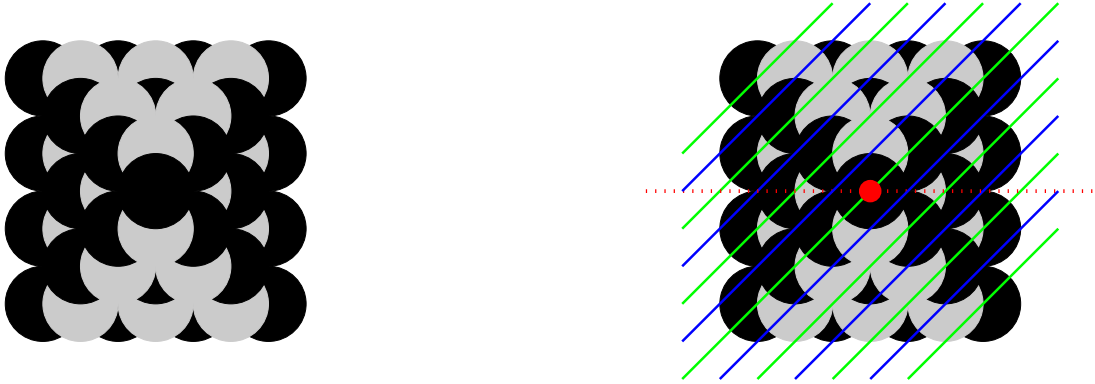}
\end{center}
\caption{Left: A 2-colored length-$1$ pyramid (empty room configuration). Right: Diagonal slicing of the pyramid. The slicing is compatible with a point reflection (indicated with the red dot) but not with a line reflection (dotted line). }
\label{ConipyramidO}
\end{figure}

As discussed in detail in \cite{Szendroi07}, the non-commutative Donaldson-Thomas partition function $\c\Zcal$ of the conifold can be inferred by counting partitions of a 2-colored pyramid of length $1$. (For the conifold, the translation from a perfect matching of the brane tilling to a pyramid partition is straight-forward, see for instance \cite{Young07}.) A 2-colored pyramid of length $N$ is an infinite stack of layers of stones, where on each layer the stones are equally colored and the coloring alternates from layer to layer. Let us label the layers by an integer $i\geq 0$. On layers $2i$ there is a rectangular array of $(i+1)(i+N)$ black stones and on layers $2i+1$ a rectangular array of $(i+1)(i+N+1)$ grey stones. (length $N$ means that there are $N$ top stones.) The arising setup is also referred to as empty room configuration and is illustrated in figure \ref{ConipyramidO} for the length $1$ case.
A pyramid partition $\pi$ in the set of length $N$ pyramid partitions $\Pcal_N$ is defined as a finite
subset of stones of the length $N$ empty room configuration such that for every stone in $\pi$ the stones on top of the stone
in the immediate layer above it (which are of different color) are also contained in $\pi$. In the length $1$ case, the partition function is given by
\beq\eqlabel{pyramidZ}
\c\Zcal(-q_0,q_1)\equiv P(q_0,q_1)=\sum_{\pi\in\Pcal_1} q_0^{|\pi|_0} q_1^{|\pi|_1}\,,
\eq
with $|\pi|_i$ the number of stones of color $i$ in $\pi$ (we take black to be $0$ and grey to be $1$). The pyramid partition function \req{pyramidZ} can be evaluated explicitly. Either via dimer shuffling \cite{Young07} or using a transfer matrix based approach \cite{Young08}. We focus on the latter way since it is the one which is most easily generalized to the orientifold case. 

Observe first that we can diagonally slice a pyramid partition, as shown in figure \ref{ConipyramidO}. The slices alternate in color and one may see each slice as a Young diagram. Let us index the slices by an integer $k\in\Z$ with $k=0$ for the central slice passing through the top stone and $k<0$ for the slices to the left and $k>0$ for the slices to the right of the central slice. We denote the 2d partition corresponding to the $k$th slice as $\pi_k$ and its transpose as $\pi'_k$. One can then show that the following interlacing conditions hold \cite{Young08}
\beq
\begin{matrix}
\pi_{2k}\succ \pi_{2k+1}\,,&\,\,\,\,\pi'_{-2k}\succ \pi'_{-(2k+1)}\,,\\
\pi'_{2k+1}\succ \pi_{2k+2}\,,&\,\,\,\,\pi_{-(2k+1)}\succ \pi_{-(2k+2)}\,.\\
\end{matrix}
\eq
These interlacing conditions combined with the alternating coloring scheme translates to the following commutator yielding the partition function \req{pyramidZ} (\cf, section \ref{transfermatrix})
\beq\eqlabel{conibasiccor}
P=\bracket{\emptypar}{\prod^\infty\left(\Gap{1}\h q_1\Gapp{1}\h q_0\right)\prod^\infty\left(\Gam{1}\h q_1\Gamp{1}\h q_0\right)}{\emptypar}\,.
\eq
Especially, the right state, as defined via \req{corrOmega}, is given by
\beq
\ket{\Omega_R}=\prod^\infty\left(\Gam{1}\h q_1\Gamp{1}\h q_0\right)\ket{\emptypar}\,.
\eq
Splitting the middle $\h q_0$ into two halves and commuting the $\h q_i$ operators on the left of it to the left state and on the right side of it to the right state via the relations \req{qcommute} yields
\beq\eqlabel{conibracket}
P=\bracket{\emptypar}{\prod^\infty\left(\Gap{u_i}\Gapp{v_i}\right)\prod^\infty\left(\Gam{v_i}\Gamp{u_i}\right)}{\emptypar}\,,
\eq
with 
\beq\eqlabel{uividef}
u_i:=q_1^{1/2}q^{i-1/2}\,,\,\,\,\,\,\,\,\, v_i:=q_1^{-1/2}q^{i-1/2}\,,  
\eq
where we defined $q:=q_0q_1$.

The usual approach to evaluate correlators like \req{conibracket} is via commuting the $\Gamma$ operators of different kind through each other such that they are annihilated by the vacuum \cite{Okounkov:2003sp}. However, as we will see below this tactic is not suitable in the orientifold case. The reason for this is that after the projection we only have one kind of operator (with kind we mean $\Gamma_+$ versus $\Gamma_-$ and similarly for the primed operators). Thus, we need to find a way to evaluate \req{conibracket} without commuting the $\Gamma_+$ through the $\Gamma_-$. Fortunately, there is a simple way to do so. 

Since the primed operators commute with the non-primed ones of the same kind, we can commute all $\Gamma_-'$ to the right and all $\Gamma_+'$ to the left. Hence, invoking the relations \req{statetoSchur}, we can express \req{conibracket} entirely in terms of Schur functions, \ie, 
\beq
P=\sum_{\lambda,\rho,\mu}s_{\lambda^t}(v)s_{\rho/\lambda}(u)s_{\rho/\mu}(v)s_{\mu^t}(u)\,.
\eq
Applying the Schur function identity \req{SchurIDc1}, we deduce
\beq
P=\prod_{i,j}\frac{1}{1-u_iv_j}\sum_{\lambda,\rho,\mu}s_{\lambda^t}(v)s_{\lambda/\rho}(v)s_{\mu/\rho}(u)s_{\mu^t}(u)\,,
\eq
which yields after invoking the identities \req{SchurIDr1} and \req{SchurIDn1}
\beq
P=\left(\prod_{i,j}\frac{1}{1-u_i v_j}\right)^2 \prod_{i,j}(1+v_iv_j)(1+u_iu_j)=M(1,q)^2 M(-q_1,q)^{-1} M(-q_1^{-1},q)^{-1}\,.
\eq
This is in agreement with the results obtained previously in \cite{Young07,Young08}.

Let us now consider the orientifold case. Note first that the diagonal slicing of the $2$-colored pyramid partition is compatible with a point-reflection but not with a line reflection, see figure \ref{ConipyramidO}. We therefore focus on the case with point-reflection. Following the definition of real non-commutative Donaldson-Thomas invariants, we have to enumerate $\T$ fixed-points which are invariant under the orientifold projection and make a choice of sign insertion $\sigma_\alpha$. This yields, 
\beq
\c \Zcal_\sigma^{real}(-q_0,q_1) \equiv R_\sigma(q_0,q_1)=\sum_{\pi\in\Scal_1}(-1)^{\frac{(1-\sigma)\tr(\pi)\pm r(\tr(\pi))}{2}}q_0^{|\pi|_0/2} q_1^{|\pi|_1/2}\,,
\eq
where $\Scal_1\subset\Pcal_1$ is the subset of $\Pcal_1$ invariant under the point reflection and $\tr(\pi)$ denotes the number of boxes on the central diagonal slice. The reason for the chosen sign will become clear below. As discussed in section \ref{transfermatrix}, the evaluation of the projected partition function boils down to evaluation of the correlator \req{Ppointref}, which reads
\beq
R_\sigma=\sum_{\lambda=\lambda^t}(-1)^{\frac{(1-\sigma)|\lambda| \pm r(\lambda)}{2}} \bracket{\lambda}{q_0^{1/2}\prod^\infty\left(\Gam{1}q_1\Gamp{1}q_0\right)}{\emptypar}\,.
\eq
Similar as in the closed case discussed above, we can express $R_\sigma$ entirely in terms of Schur functions, namely
\beq
\begin{split}
R_\sigma&=\sum_{\lambda=\lambda^t}(-1)^{\frac{(1-\sigma)|\lambda|\pm r(\lambda)}{2}}  \sum_{\mu}s_{\lambda/\mu}(v)s_{\mu^t}(u)  \\
&=\sum_{\lambda=\lambda^t}(-1)^{\frac{|\lambda|-\sigma r(\lambda)}{2}} \sum_{\mu}s_{\lambda/\mu}(\ii v)s_{\mu^t}(\ii u)  \,.
\end{split}
\eq
Using the identities \req{SchurIDo2} and \req{SchurIDr2} we infer
\beq\eqlabel{Rconiprefinal}
R_\pm(u_i,v_i)=\prod_{i}(1\pm \ii v_i)\prod_{i<j}(1+v_iv_j)\prod_{i,j}\frac{1}{1-u_i v_j} \sum_{\lambda=\lambda^t}  (-1)^{\frac{|\lambda|\mp r(\lambda)}{2}} s_\lambda(\ii u)\,.
\eq
Invoking the Schur function identity \req{SchurIDo1}, we arrive at
\beq\eqlabel{realconifinalP}
R_\pm=M(1,q)\, M_\pm^{real}(-q_1,q)^{-1}\,M_\pm^{real}(-q_1^{-1},q)^{-1}\,.
\eq
Using the same reparameterization as in \cite{Szendroi07}, \ie,
\beq
 Q \rightarrow -q_1,\,\,\, q\rightarrow -q_0q_1\,,
\eq
we deduce
\beq\eqlabel{NCDTConiFinal}
\c \Zcal^{real}_\pm(Q,q)=M(1,q)\, M_\pm^{real}(Q,-q)^{-1}M_\pm^{real}(Q^{-1},-q)^{-1}\,,
\eq
with $\c \Zcal^{real}_\pm(Q\rightarrow -q_1,q\rightarrow -q_0q_1)=\c\Zcal_\pm^{real}(q_0,q_1)$. This confirms the prediction made on general grounds in the introduction. Also note that we infer from \req{realconifinalP} that the degree $0$ contribution is indeed $M(1,q)$, confirming the discussion of section \ref{d0contribution}.

Now it is clear why we choose the $\sigma_\alpha$ sign in \req{realNCDTinv} to be the $r$-type sign for the point-reflection conifold example. This sign is necessary to be able to transform the Schur function summation over $\lambda=\lambda^t$ into the expected product form via the identities \req{SchurIDo1} and \req{SchurIDo2}.

\begin{figure}
\begin{center}
\includegraphics[scale=0.5]{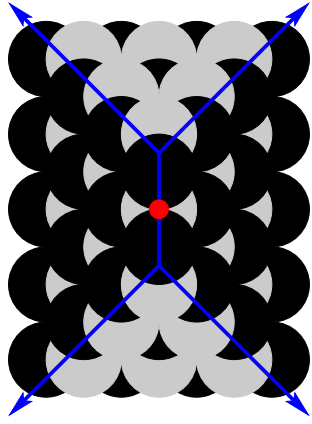}
\end{center}
\caption{Empty room configuration of a length $2$ pyramid. The big red dot marks the origin of the point reflection symmetry and the blue lines indicate the web diagram.}
\label{ConipyramidOl2}
\end{figure}

We can consider as well pyramids of length $N$ with partition function denoted as $P_N$. For example, the empty room configuration of length $2$ is shown in figure \ref{ConipyramidOl2}.
The physical interpretation (without orientifold) of the corresponding partition functions has been given in \cite{Chuang:2008aw} (see also \cite{Nagao08}) there it was found that they correspond to counting D6-D2-D0 bound states in different chambers, \ie, in quiver language for different choices of stability parameters. Thus, we have that up to reparameterization
\beq
Z_{BPS}^{(N)}\equiv\c\Zcal_N(-q_0,q_1)= P_N(q_0,q_1)\,.
\eq

The partition function $P_{N}$ can be obtained via blowing up the length $1$ pyramid as sketched in section \ref{transfermatrix}, \ie, via inserting the operator $\h\Wcal^{N-1}$ into \req{conibasiccor}. We know that $\h \Wcal^{N-1}$ takes the form \cite{Sulkowski:2009rw}
\beq
\h \Wcal^{N-1}=\left[\Gam{1}\h q_1\Gapp{1}\h q_0\right]^{N-1}\,.
\eq
In order to determine the real partition function $R_{N,\sigma}$ we split $\h \Wcal^{N-1}$ into a left and right part as in \req{Wsplit}. As is clear from figures \ref{ConipyramidO} and \ref{ConipyramidOl2}, the splitting is $N$ dependent. For $N-1$ even the origin of the point reflection sits on a black stone, while for $N-1$ odd it sits on a grey stone. Let us start with the $N-1$ even case.
We have
\beq
\h\Wcal_R=\h q_0^{1/2}\left[\Gam{1}\h q_1\Gapp{1}\h q_0 \right]^{(N-1)/2}\,.
\eq
Commuting the $\h q_i$ operators to the right, the $\Gamma'_+$ operators to the left, and defining $\h x=(\h q_0\h q_1)^{(N-1)/2}$ yields
\beq
\h\Wcal_R=\prod_{i\leq j}^{(N-1)/2}\left(\frac{1}{1+v_iu_j^{-1}}\right)\prod_{i=1}^{(N-1)/2}\left[\Gapp{u_i^{-1}}\right] \prod_{i=1}^{(N-1)/2}\left[\Gam{v_i}\right]\h x \h q_0^{1/2}  \,,
\eq
where we used \req{Gcommute} and $u_i,v_i$ is defined as in \req{uividef}. Thus, we have to evaluate
\beq\eqlabel{RconiNodd0}
\begin{split}
R_{N\,{\rm odd},\sigma}&=\sum_\lambda (-1)^{\frac{(1-\sigma)|\lambda|\pm r(\lambda)}{2}} \bracket{\lambda}{\h\Wcal_R}{\Omega_R}\\
&=\prod_{i\leq j}^{(N-1)/2}\left(\frac{1}{1+v_iu_j^{-1}}\right)\\
&\times\sum_\lambda  (-1)^{\frac{(1-\sigma)|\lambda|\pm r(\lambda)}{2}} \bracket{\lambda}{\prod_{i=1}^{(N-1)/2}\left[\Gapp{u_i^{-1}}\right] \prod_{i=1}^\infty\left[\Gam{v_i}\Gamp{x\cdot u_i}\right]    }{\emptypar}\,.
\end{split}
\eq
The evaluation of \req{RconiNodd0} is elementary. We will give the details in appendix \ref{LNconicalc}.

In the $N-1$ odd case, we have
\beq
\h\Wcal_R=\h q_1^{1/2}\Gapp{1}\h q_0 \left[\Gam{1}\h q_1\Gapp{1}\h q_0\right]^{N/2-1}\,,
\eq
which can be rewritten similar as in the $N-1$ even case as
\beq
\h\Wcal_R=\prod_{i\leq j}^{N/2-1}\frac{1}{1+v_iu_j^{-1}}\prod_{i=1}^{N/2}\left[\Gapp{q^{1/2}u_i^{-1}}\right] \prod_{i=1}^{N/2-1}\left[\Gam{q_1^{i-1/2}q_0^{i}}\right]  \h x \h q_0^{1/2}\,.
\eq
We infer
\beq\eqlabel{RconiNeven0}
\begin{split}
R_{N\,{\rm even},\sigma}&=\prod_{i\leq j}^{N/2-1}\frac{1}{1+v_iu_j^{-1}}\\
&\times \sum_\lambda (-1)^{\frac{(1-\sigma)|\lambda|\pm r(\lambda)}{2}} \bracket{\lambda}{\prod_{i=1}^{N/2}\left[\Gapp{q^{1/2}u_i^{-1}}\right] \prod_{i=1}^\infty\left[\Gam{q^{1/2}v_i}\Gamp{x\cdot u_i}\right]}{\emptypar}\,.
\end{split}
\eq
We again perform the evaluation of the correlator in appendix \ref{LNconicalc}. 

From the calculations in the appendix we deduce that
\beq\eqlabel{Rconifinal}
R_{N,\sigma}=M(1,q) \,M^{real}_\sigma(-q_1q^{N-1},q)^{-1}\, M_{N,\sigma}^{real}(-q_1^{-1}q^{-(N-1)},q)^{-1}\,,
\eq
with $M_{N,\sigma}^{real}(x,q)$ as defined in \req{MNrealdef}. With the redefinition of parameters
\beq\eqlabel{Qredef}
Q:=-q^{N-1}q_1\,,
\eq
and $q\rightarrow -q$, we arrive at
\beq
\c\Zcal_{N,\sigma}(Q,q)=M(1,-q)\,M^{real}_\sigma(Q,-q)^{-1}\, M_{N,\sigma}^{real}(Q^{-1},-q)^{-1}\,.
\eq
Especially, for $N=1$ we recover \req{NCDTConiFinal}, while for $N\rightarrow\infty$ we approach \req{ConirealDTZ}. 
Defining
\beq
\c \Zcal'_{N,\sigma}=\frac{\c\Zcal^{real}_{N,\sigma}}{\c\Zcal^{1/2}_{N,\sigma}}\,,
\eq
we infer that
\beq
{Z'}^{(N)}_{BPS}\equiv \c \Zcal'_{N,\pm}(Q,q)=\Zcal'_\pm(Q,q)\,M'_{N,\pm}(Q^{-1},-q)^{-1}=\Zcal'_\pm(Q,q)\prod_{i>\lfloor N/2\rfloor}^\infty\Wcal_i  \,,
\eq
with
\beq\eqlabel{ConiWired}
\Wcal_i=\left(\frac{1\mp Q^{-1/2}(-q)^{i-1/2} }{1\pm Q^{-1/2}(-q)^{i-1/2}}\right)^{1/2}\,,
\eq
where we used \req{MNreduced2} and \req{MNreduced1}. 

\subsection{Example 2: Non-commutative $\C^2/ \Z_2\times \C$}
\label{c2z2Nsec}
\begin{figure}
\begin{center}
\includegraphics[scale=0.33]{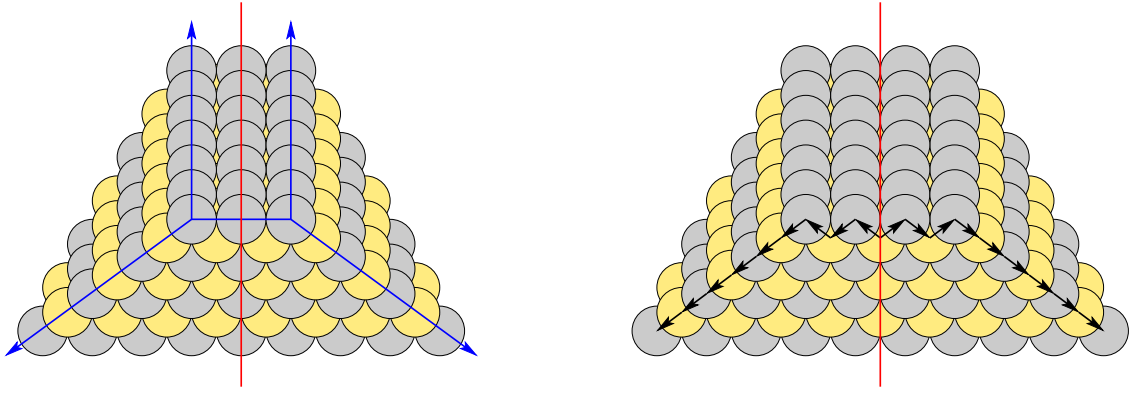}
\end{center}
\caption{Left: Crystal setup for $N=3$ with indicated web diagram. Right: Setup for $N=4$ with indicated order of $\Gamma$ operators.}
\label{C2Z2symblowup}
\end{figure}

The crystal model associated to the non-commutative resolution of $\C^2/ \Z_2\times \C$ has been discussed in \cite{Sulkowski:2009rw} and corresponds to a 2-colored triangular stack of stones. Note that the partition function of this crystal is equivalent to the partition function of the 2-colored plane partition discussed in section \ref{orbifoldsec} (the same holds for all $\C^2/ \Z_n\times \C$). Hence, the orbifold partition function is equivalent to the non-commutative one. However, since we will need below some parts of the derivation of the partition function, we will briefly rederive the partition function here in a more economical fashion than in the more general derivation done in appendix \ref{C2ZNxCcalc}. That is, from \req{C2ZnCcor} and \req{statetoSchur} we infer
\beq
\c\Zcal(-q_0,q_1)\equiv R_\sigma(q_0,q_1)=\sum_{\lambda} (-1)^{\frac{(1-\sigma)|\lambda|}{2}}\sum_\mu s_{\lambda/\mu}(v)s_{\mu}(u)\,.
\eq  
Using \req{SchurIDo4}, \req{SchurIDr2} and \req{SchurIDo3} we deduce
\beq\eqlabel{RC2Z2final}
\begin{split}
R_\pm&=\prod_i\frac{1}{1\pm v_i} \prod_{i<j}\frac{1}{1-v_iv_j}\prod_{i,j}\frac{1}{1-v_iu_j}\sum_\lambda (-1)^{\frac{(1\pm 1)|\lambda|}{2}}s_\lambda(u) \\
&=M(1,q)\, M^{real}_\pm(q_1,q)\, M^{real}_\pm(q_1^{-1},q)\,.
\end{split}
\eq
Let us now consider the blow up of the crystal via inserting powers of a wall crossing operator $\h\Wcal$ (see section \ref{transfermatrix}) given by
\beq
\h\Wcal=\left[\Gam{1}\h q_1\Gap{1}\h q_0 \right]\,.
\eq
For illustration, some blown up stack of stones are shown in figure \ref{C2Z2symblowup}.
The line reflection of the crystal setup sits either on a 2d partition composed of grey stones (operator $\h q_0$) for $N-1$ even or on a 2d partition of yellow stones (operator $\h q_1$) for $N-1$ odd. Let us start with the $N-1$ even case. We split $\h\Wcal^{N-1}$ as in \req{Wsplit} and we have (very similar to the conifold discussed in section \ref{conisec}, essentially, the difference is just a substitution $\Gamma_+'\rightarrow\Gamma_+$)
\beq
\h\Wcal_R= \h q_0^{1/2}\left[\Gam{1}\h q_1\Gap{1}\h q_0 \right]^{(N-1)/2}\,.
\eq
With the basic right state $\ket{\Omega_R}$, which can be inferred from \req{C2Znbasiccorr} to be given by
\beq
\ket{\Omega_R}=\prod^\infty\left[\Gam{1}\h q_1\Gam{1} \h q_0\right]\ket{\emptypar}\,,
\eq
the evaluation of
\beq
R_{N\,{\rm odd},\sigma}=\sum_\lambda (-1)^{\frac{(1-\sigma)|\lambda|}{2}}\bracket{\lambda}{\h\Wcal_R}{\Omega_R}\,,
\eq
boils down to evaluating
\beq\eqlabel{C2Z2lNoddcorr}
\begin{split}
R_{N\,{\rm odd},\sigma}&=\prod_{i\leq j}^{(N-1)/2}\left( 1-v_i u_j^{-1}\right)\\&\times\sum_\lambda(-1)^{\frac{(1-\sigma)|\lambda|}{2}}\bracket{\lambda}{\prod_{i=1}^{(N-1)/2}\left[\Gap{u_i^{-1}}\right]\prod_{i=1}^{\infty}\left[\Gam{x \cdot u_i}\Gam{v_i}\right]}{\emptypar}\,.\\
\end{split}
\eq
The details of which will be given in appendix \ref{C2Z2correv}.

For $N-1$ odd we have instead the wall crossing operator
\beq
\h\Wcal_R=\h q_1^{1/2}\Gap{1}\h q_0 \left[\Gam{1}\h q_1\Gap{1}\h q_0\right]^{N/2-1}\,,
\eq
which is again almost the same as in the conifold case. We deduce
\beq\eqlabel{C2Z2lNevencorr}
\begin{split}
R_{N\,{\rm even},\sigma}&=\prod_{i\leq j}^{N/2-1}\left( 1-v_i u_j^{-1}\right)\\&\times\sum_\lambda(-1)^{\frac{(1-\sigma)|\lambda|}{2}}\bracket{\lambda}{\prod_{i=1}^{N/2}\left[\Gap{q^{1/2}u_i^{-1}}\right]\prod_{i=1}^{\infty}\left[\Gam{x \cdot u_i}\Gam{q^{1/2}v_i}\right]}{\emptypar}\,.\\
\end{split}
\eq
The evaluation can be found in appendix \ref{C2Z2correv}. The final result is given by
\beq\eqlabel{C2Z2lengthNfinal}
R_{N,\sigma}=M(1,q)\,M^{real}_\sigma(q_1 q^{N-1},q)\,M^{real}_{N,\sigma}(q_1^{-1} q^{-(N-1)},q)\,,
\eq
from which we obtain by the redefinition of variables $Q:=q^{N-1}q_1$ and $q\rightarrow -q$, 
\beq
\c\Zcal_{N,\sigma}(Q,q)=M(1,-q)\,M^{real}_\sigma(Q,-q)\,M^{real}_{N,\sigma}(Q^{-1},-q)\,.
\eq
For $N=1$ we recover \req{RC2Z2final} and for $N\rightarrow\infty$ we approach \req{ZC2Z2final}. The reduced wall crossing factor $\Wcal'$ can be inferred as in the previous section, and is identical to \req{ConiWired}.

\section{Conclusion}
\label{concl}

In this work, we have initiated the study of wall crossing phenomena in orientifolds of local toric backgrounds. For this purpose, we gave a definition of real Donaldson-Thomas invariants at several points in K\"ahler moduli space, \ie, at the large volume, orbifold and non-commutative point. Especially, the Donaldson-Thomas formulation allowed us to confirm and refine the constant map conjecture to the real topological string partition function made in \cite{Krefl:2009mw}. In detail, the definitions are based on localization with respect to the subtorus $\T'\subset\T$ surviving the orientifold projection plus certain sign insertions. Though, the made definitions are on solid ground, it would be desirable to formulate real Donaldson-Thomas invariants in a more rigirous, respectively direct way, than we did in sections \ref{realDTinvariants}, \ref{orbidef} and \ref{defNCDT}. It would be very interesting to see if one can derive the signs which we inserted in an ad hoc fashion from first principles. For example, at the large volume point via carefully performing the localization calculation of \cite{MNOPI} in a real setting or at the non-commutative point via deriving the quiver whose path algebra yields the non-commutative resolution of the orientifolded background (in the quotient). Presumably, in gauge theory language the corresponding quiver might involve enhanced gauge groups and 2-tensor representations and it would be interesting to see how the sign insertions translate to this language.

We conjectured that, at least for models without compact divisors, the (properly reparameterized) real Donaldson-Thomas partition function at the orbifold, respectively non-commutative point in K\"ahler moduli space is related to the real Donaldson-Thomas partition function at the large volume point via the relation
\beq\eqlabel{Zfinalconjec}
\c \Zcal^{real}(Q,q)=\t \Zcal^{real}(Q^{-1},q)\,\t\Zcal^{real}(Q,q)\,,
\eq
\ie, the wall crossing factor $\Wcal$ is simply given by $\Wcal=\t\Zcal^{real}(Q^{-1},q)$.  Explicit confirmation for this proposal has been found at hand of the conifold and the orbifold $\C^2/\Z_n\times\C$, for which we explicitly computed $\c\Zcal^{real}$ combinatorially. (Note that these calculations also confirm the predicted degree $0$ contribution.) 

Especially, for models with $n^{(g>0)}_{\vec d}=0$, the basic building block of the reduced wall crossing factor $\Wcal'$ (defined in equation \req{Wrealsplit}) reads
\beq\eqlabel{Wreducedfinalc}
\Wcal'_i= \prod_{\vec d_a, \vec d_b} \left(\frac{1-Q_{\vec d_a}^{-1/2}Q^{-1}_{\vec d_b}q^{i-1/2}}{1+Q_{\vec d_a}^{-1/2}Q^{-1}_{\vec d_b}q^{i-1/2}}\right)^{\pm N^{(-1)}_{\vec d_a, \vec d_b}/2}\,.
\eq 
This factor is qualitatively similar to the pure open wall crossing factor derived in \cite{Aganagic:2009cg}, but includes in addition a contribution arising from the unoriented sector.

It would be nice to obtain an understanding of the relations \req{Zfinalconjec} and \req{Wreducedfinalc} from a BPS state counting point of view. This could be achieved either from a M-theory perspective, perhaps similar as in \cite{Aganagic:2009kf,Aganagic:2009cg}, or from a supergravity point of view following \cite{Denef:2007vg,Jafferis:2008uf}. For the M-theory approach, it would be useful to first give a M-theory derivation of the real Gopakumar-Vafa expansion \req{realexpansion}, which would also be of more general interest. Namely, it would be interesting to understand real Gopakumar-Vafa invariants from a M-theory perspective along the lines of \cite{Gopakumar:1998ii,Gopakumar:1998jq,Katz:1999xq,Ooguri:1999bv}. In contrast, for the supergravity approach one should first derive an orientifold version of the semi-primitive wall crossing formula of \cite{Denef:2007vg}, which is an interesting problem of its own. Turned around, \req{Zfinalconjec} and \req{Wreducedfinalc} give some hints how such a formula should look like.

Finally, one should note that our orientifold action (and hence the real topological string) is different from the orientifold action used in \cite{Denef:2009ja}, where an orientifold version of the OSV conjecture \cite{Ooguri:2004zv} was proposed. 
It would be interesting to clarify the similarities/differences between the two cases. One possible approach which might lead to further insight into the orientifold version of the OSV conjecture would be to reconsider \cite{Aganagic:2004js,Aganagic:2005dh} for a general Riemannian surface, \ie, with boundaries and crosscaps.

We are looking forward to further research on these matters.

\acknowledgments{
I would like to thank M. Yamazaki for initial collaboration, F. Denef, K. Hori and J. Walcher for related communications/discussions and Caltech, YITP Kyoto and MSRI Berkeley for kind hospitality during part of this project. My work was supported by the WPI initiative by MEXT of Japan.
}

\newpage
\appendix

\section{Generalized MacMohan functions}
\label{appB}
For the readers convenience, we recall in this appendix the definitions of the generalized MacMohan functions, which play a major role in the main text. The generalized MacMohan function is the generating function for weighted 3d partitions (also known as plane partitions) and the most compact expression is given by
\beq\eqlabel{appBMh}
M(x,q):=\sum_{\pi \in \Pcal}x^{\tr(\pi)}q^{|\pi|} = \prod_{n=1}^\infty\frac{1}{(1-x q^n)^n}\,,
\eq
where $\Pcal$ is the set of plane partitions, $|\pi|$ denotes the number of boxes in $\pi$ and $\tr(\pi)$ the number of boxes on the diagonal. It is easy to see that we can express $M(x,q)$ via the following double product
\beq\eqlabel{appBMhdouble}
\begin{split}
M(x,q)=\prod_{i,j}^\infty\frac{1}{(1-xq^{i+j-1})}&=\prod_{i\leq j}^\infty\frac{1}{(1-xq^{i+j-1})}\prod_{i>j}^\infty\frac{1}{(1-xq^{i+j-1})}\\
&=\prod_{i=1}^\infty \frac{1}{(1-xq^{2i-1})}\left(\prod_{i<j}^\infty\frac{1}{(1-xq^{i+j-1})}\right)^2\,.
\end{split}
\eq
We define the function $M_N(x,q)$ as the product in \req{appBMh} starting from $n=N$, \ie,
\beq\eqlabel{NMacMohan}
M_N(x,q):=\prod_{n=N}^\infty \frac{1}{(1-xq^n)^n}\,.
\eq
Note that one can express $M_N(x,q)$ as
\beq\eqlabel{Mn1}
\begin{split}
M_N(x,q)&=\prod_{i,j}^\infty\frac{1}{1-xq^{N-1}q^{i+j-1}}\prod_{i=1}^\infty\frac{1}{(1-x q^{N-1}q^{i})^{N-1}}\\
&=M(x q^{N-1},q)\prod_{i=1}^\infty\frac{1}{(1-x q^{N-1}q^{i})^{N-1}}
\,,
\end{split}
\eq
or, alternatively, as
\beq\eqlabel{Mn2}
\begin{split}
M_N(x,q)&=\prod_{i,j}^\infty\frac{1}{1-xq^{N}q^{i+j-1}}\prod_{i=1}^\infty\frac{1}{(1-x q^{N-1}q^{i})^{N}}\\
&=M(x q^{N},q)\prod_{i=1}^\infty\frac{1}{(1-x q^{N-1}q^{i})^{N}}\,,
\end{split}
\eq
where the latter expression can be obtained from \req{Mn1} by rearranging factors. Especially, we observe that $M_1(x,q)=M(x,q)$ and $M_\infty(x,q)=1$.

The generalized real MacMohan function is defined to be the generating function of weighted symmetric plane partitions 
\beq\eqlabel{realMacMohanstandard}
M^{real}_\pm(x,q):=\sum_{\pi\in\Scal}(\mp 1)^{\tr(\pi)} x^{\tr(\pi)/2}q^{|\pi|/2}=\prod_{n=1}^\infty\frac{1}{(1\pm x^{1/2}q^{n-1/2})}\prod_{n=1}^\infty\frac{1}{(1-xq^{n})^{\lfloor n/2\rfloor} }\,,
\eq
where $\Scal\subset\Pcal$ is the subset of symmetric plane partitions. The additional weighting $x\rightarrow x^{1/2}$ and $q\rightarrow q^{1/2}$ occurs because we quotient by the symmetry, \ie, only half of the boxes are in the quotient. Especially, the boxes on the diagonal are cut into half-boxes and therefore each box on the diagonal contributes only by a factor of $1/2$.

The last factor in \req{realMacMohanstandard} can be rewritten as a double product, \ie, 
\beq\eqlabel{appBrealMhdouble}
M^{real}_\pm(x,q)=\prod_{i=1}^\infty\frac{1}{(1\pm x^{1/2}q^{i-1/2})}\prod_{i<j}^\infty \frac{1}{(1-xq^{i+j-1})}\,.
\eq
Similar as in \req{NMacMohan}, we define the function
\beq\eqlabel{MNrealdef}
M^{real}_{N,\pm}(x,q)=
\prod_{n>\lfloor N/2\rfloor}^\infty\frac{1}{(1\pm x^{1/2}q^{n-1/2})}\prod_{n=N}^\infty\frac{1}{(1-xq^{n})^{\lfloor n/2\rfloor} }\,,
\eq
which can be rewritten as
\beq\eqlabel{MNrealfull}
\begin{split}
M^{real}_{N,\pm}(x,q)&=\prod_{i=1}^\infty\frac{1}{(1\pm x^{1/2}q^{\lfloor N/2\rfloor }q^{i-1/2})}\prod_{i<j}^\infty\frac{1}{(1-xq^{2\lfloor N/2\rfloor}q^{i+j-1})}\prod_{i=1}^\infty\frac{1}{(1-x q^{N-1}q^i)^{\lfloor N/2\rfloor}}\\
&=M^{real}_\pm(xq^{2\lfloor N/2\rfloor},q) \prod_{i=1}^\infty\frac{1}{(1-x q^{N-1}q^i)^{\lfloor N/2\rfloor}} \,.
\end{split}
\eq
We observe that $M^{real}_{1,\pm}(x,q)=M^{real}_\pm(x,q)$ and $M^{real}_{\infty,\pm}(x,q)=1$.

Finally, the reduced generalized MacMohan function $M'_\pm(x,q)$ is defined via 
\beq\eqlabel{appBreducedrealMac}
M'_\pm(x,q):=\frac{M^{real}_\pm(x,q)}{M(x,q)^{1/2}}\,.
\eq
Using \req{appBMhdouble} and \req{appBrealMhdouble}, we immediately deduce that
\beq\eqlabel{appBreducedrealMac}
M'_\pm(x,q)=\prod_{n=1}^\infty\frac{(1- x q^{2n-1})^{1/2}}{1\pm x^{1/2} q^{n-1/2}}=\prod_{n=1}^\infty\left(\frac{1\mp x^{1/2}q^{n-1/2}}{1\pm x^{1/2}q^{n-1/2}}\right)^{1/2}\,.
\eq
Let us define
\beq\eqlabel{MNreduced1}
\begin{split}
M'_{N,\pm}(x,q)&:=\prod_{n>\lfloor N/2\rfloor}^\infty \left(\frac{1\mp x^{1/2}q^{n-1/2}}{1\pm x^{1/2}q^{n-1/2}}\right)^{1/2}=\prod_{n=1}^\infty \left(\frac{1\mp x^{1/2}q^{\lfloor N/2\rfloor}q^{n-1/2}}{1\pm x^{1/2}q^{\lfloor N/2\rfloor}q^{n-1/2}}\right)^{1/2}\\
&=M'_{\pm}(xq^{2\lfloor N/2\rfloor},q)\,.
\end{split}
\eq
For $N=1$ we recover $\req{appBreducedrealMac}$ while for $N\rightarrow\infty$ we have $M'_{\infty,\pm}(x,q)=1$. Using for $N$ odd \req{Mn1} and for $N$ even \req{Mn2}, we deduce that 
\beq\eqlabel{MNreduced2}
M'_{N,\pm}(x,q)=\frac{M^{real}_{N,\pm}(x,q)}{M_{N}(x,q)^{1/2}}\,,
\eq
holds, consistent with \req{appBreducedrealMac}.

\section{Schur function identities}
\label{Schur}

In this appendix we collect Schur function identities which are used in the main text. Derivations and detailed discussions can be found for instance in \cite{Stanley,MacDonald}.

Firstly, we have the well-known orthogonality relations of Schur functions
\beq\eqlabel{SchurIDn1}
\sum_\lambda s_\lambda(x) s_\lambda(y)=\prod_{i,j}\frac{1}{1-x_i y_j}\,,
\eq
\beq\eqlabel{SchurIDn2}
\sum_\lambda s_{\lambda^t}(x) s_\lambda(y)=\prod_{i,j}(1+x_i y_j)\,.\\
\eq

If we replace in \req{SchurIDn1} and \req{SchurIDn2} the Schur function $s_\lambda(y)$ by a skew Schur function $s_{\lambda/\rho}(y)$, the identities change to
\beq\eqlabel{SchurIDr2}
\sum_\lambda s_{\lambda}(x)s_{\lambda/\rho}(y)=s_\rho(x) \prod_{i,j}\frac{1}{(1-x_i y_j)}\,,
\eq 

\beq\eqlabel{SchurIDr1}
\sum_\lambda s_{\lambda^t}(x)s_{\lambda/\rho}(y)=s_\rho(x) \prod_{i,j}(1+x_i y_j)\,.
\eq

Replacing also the remaining Schur functions in \req{SchurIDr2} and \req{SchurIDr1} by skew Schur functions, the relations get modified to
\beq\eqlabel{SchurIDc1}
\sum_\rho s_{\rho/\lambda}(x)s_{\rho/\mu}(y)=\prod_{i,j}\frac{1}{1-x_iy_j}\sum_\rho s_{\lambda/\rho}(y)s_{\mu/\rho}(x)\,.
\eq
\beq\eqlabel{SchurIDc2}
\sum_\rho s_{\rho^t/\lambda}(x)s_{\rho/\mu}(y)=\prod_{i,j}(1+x_iy_j)\sum_\rho s_{\lambda^t/\rho}(y)s_{\mu^t/\rho^t}(x)\,.
\eq

Especially, for $x=q^{\rho}$ and $y=Qq^{\rho}$ the products in the above identities yield generalized MacMohan functions $M(Q,q)$, respectively $M(-Q,q)^{-1}$,  as is apparent from \req{appBMh}.

In the real case, the analog of the orthogonality relations \req{SchurIDn1} and \req{SchurIDn2} may be seen in the following relation
\beq\eqlabel{SchurIDo3}
\sum_\lambda   (\mp 1)^{|\lambda|}s_\lambda(x)=\prod_i\frac{1}{1\pm x_i}\prod_{i<j}\frac{1}{1-x_ix_j}\,.
\eq
For $x=Q^{1/2}q^{\rho}$ this identity yields a generalized real MacMohan function $M^{real}_\pm(Q,q)$, as is clear from the definition given in \req{appBrealMhdouble}.

Similarly, the real analog of \req{SchurIDc1} and \req{SchurIDc2} is the relation
\beq\eqlabel{SchurIDo4}
\sum_{\rho}(\mp 1)^{|\rho|-|\lambda|}s_{\rho/\lambda}(x)=\prod_i\frac{1}{1\pm x_i}\prod_{i<j}\frac{1}{1-x_ix_j}\sum_\rho(\mp 1)^{|\lambda|-|\rho|} s_{\lambda/\rho}(x)\,.
\eq

Finally, the following two identities are particularly useful in the real case with point-reflection

\beq\eqlabel{SchurIDo1}
\sum_{\lambda=\lambda^t} (-1)^{\frac{|\lambda|\mp r(\lambda)}{2}}s_\lambda(x)=\prod_{i=1}(1\pm x_i)\prod_{i<j}(1-x_i x_j)\,,
\eq

\beq\eqlabel{SchurIDo2}
\sum_{\lambda=\lambda^t} (-1)^{\frac{|\lambda|\mp r(\lambda)}{2}} s_{\lambda/\mu}(x)=\prod_{i=1}(1\pm x_i)\prod_{i<j}(1-x_ix_j)\sum_{\lambda=\lambda^t} (-1)^{\frac{|\lambda|\mp r(\lambda)}{2}} s_{\mu^t/\lambda}(-x)\,.
\eq
Note that for $x=Q^{1/2}q^{\rho}$ the occuring products yield $M^{real}_\pm(Q,q)^{-1}$.

\section{Evaluation of correlators}
\label{appC}
In this appendix we give details about the evaluation of certain correlators occuring in the examples discussed in sections \ref{orbifoldsec}, \ref{conisec} and \ref{c2z2Nsec}. The calculations are elementary, however for completeness we include them here.

\subsection{Derivation of real partition function of $\C^2/\Z_n\times\C$}

\label{C2ZNxCcalc}
In this subsection we evaluate the correlator \req{C2ZnCcor}. Invoking the relations \req{statetoSchur}, we express the correlator \req{C2ZnCcor} in terms of Schur functions as follows
\beq
\begin{split}
P^{real}_{n,\sigma}=\sum_\lambda &(-1)^{\frac{(1-\sigma)|\lambda|}{2}}\\
&\times s_\lambda\left(q_1^{-1/2}\dots q_{n-1}^{-1/2}q^{1/2},q_1^{1/2}q_2^{-1/2}\dots q_{n-1}^{-1/2} q^{1/2},\dots,q_1^{1/2}\dots q_{n-1}^{1/2} q^{1/2},\right.\\
&\left. q_1^{-1/2}\dots q_{n-1}^{-1/2}q^{3/2},q_1^{1/2}q_2^{-1/2}\dots q_{n-1}^{-1/2} q^{3/2},\dots,q_1^{1/2}\dots q_{n-1}^{1/2} q^{3/2},\dots \right)\,.
\end{split}
\eq
Let us distinguish between the case with $n$ odd and $n$ even. 

\paragraph{$n$ odd}

For $n$ odd, only $q_0$ is fixed, hence, after identifying colors, 
\beq\eqlabel{Pnodd}
\begin{split}
P^{real}_{n\,{\rm odd},\sigma}=&\sum_\mu (-1)^{\frac{(1-\sigma)|\lambda|}{2}}\\
&\times s_\mu\left(q^{-1}_{[1,\frac{n-1}{2}]}q^{1/2},q^{-1}_{[2,\frac{n-1}{2}]}q^{1/2},\dots,q^{1/2},q_{\frac{n-1}{2}}  q^{1/2},q_{[\frac{n-1}{2},\frac{n-3}{2}]}q^{1/2},\dots,q_{[\frac{n-1}{2},1]} q^{1/2}, \right.\\
&\left. q^{-1}_{[1,\frac{n-1}{2}]}q^{3/2},q^{-1}_{[2,\frac{n-1}{2}]}q^{3/2},\dots,q^{3/2},q_{\frac{n-1}{2}} q^{3/2},q_{[\frac{n-1}{2},\frac{n-3}{2}]} q^{3/2},\dots,q_{[\frac{n-1}{2},1]} q^{3/2},\right.\\
&\left.\phantom{\sum} \dots\right )\,,
\end{split}
\eq
where we defined $q_{[i,j]}$ similar as in \req{Qkomdef} and $q$ is now given by $q=q_0^{1/2}q^2_{[1,\frac{n-1}{2}]}$.

We solve  \req{Pnodd} via using the Schur function identity \req{SchurIDo3}. The first factor of \req{SchurIDo3} yields
\beq\label{firstpart}
\prod_{i=1}^\infty \prod_{j=1}^{(n-1)/2} \left(\frac{1}{1\pm q_{[j,\frac{n-1}{2}]}q^{i-1/2}}\right) \left(\frac{1}{1\pm q^{-1}_{[j,\frac{n-1}{2}]}q^{i-1/2}} \right)\left(\frac{1}{1\pm q^{i-1/2}}\right)\,.
\eq
In order to evaluate the second factor of \req{SchurIDo3}, let us define $K_0=\frac{n+1}{2}$, $K_1=\{1,\dots,\frac{n-1}{2}\}$, $K_2=\{\frac{n+3}{2},\dots,n\}$ and let $K_i^m$ denote the $m$-th element of the set $K_i$. We define indices $k_i^m$ and ${k'}_i^m$ running over $\{K_i^m, 2K_i^m,3K_i^m,\dots\}$ and reexpress the product in the second factor of  \req{SchurIDo3} in these variables, \ie, 
\beq\label{secondpart}
\begin{split}
\prod_{i<j}&=\left(\prod_{k^1_0<{k'}_0^1}\right)\left(\prod_{j=1}^{(n-1)/2} \prod_{k^1_0<k_1^j}\right)\left(\prod_{j=1}^{(n-1)/2} \prod_{k_0^1<k_2^j}\right)\\
&\times\left(\prod_{j=1}^{(n-1)/2} \prod_{k_1^j<k_0^1}\right)\left(\prod_{i,j=1}^{(n-1)/2} \prod_{k_1^i<k_2^j}\right)\left(\prod_{1\leq i,j\leq\frac{n-1}{2}} \prod_{k_1^i<k_1^j}\right)\\
&\times\left(\prod_{j=1}^{(n-1)/2} \prod_{k_2^j<k_0^1}\right)\left(\prod_{i,j}^{(n-1)/2} \prod_{k_2^i<k_1^j}\right)\left(\prod_{1\leq i,j\leq\frac{n-1}{2}} \prod_{k_2^i<k_2^j}\right)\,.
\end{split}
\eq
We have to evaluate nine factors. The first factor
\beq
\prod_{k^1_0<{k'}_0^1}\frac{1}{1-x_{k_0^1} x_{{k'}_0^1}}=\prod_{i<j}\frac{1}{1-q^{i+j-1}}\,,
\eq
combines with the last factor of \req{firstpart} to a real MacMohan $M_\sigma^{real}(1,q)$ which we associate with the constant map contribution. 

Let us split the sixth factor into two factors with $i<j$ and $i\geq j$, \ie, 
\beq\eqlabel{6factorA}
\prod_{1\leq i< j\leq (n-1)/2)}\prod_{k_1^i<k_1^j}\frac{1}{1-x_{k_1^i}x_{k_1^j}}=\prod_{k< l}^{(n-1)/2} \prod_{i\leq j} \frac{1}{1-q^{-1}_{[k,\frac{n-1}{2}]}q^{-1}_{[\frac{n-1}{2},l]}q^{i+j-1}}\,,
\eq
and
\beq\eqlabel{6factorB}
\begin{split}
&\prod_{1\leq j\leq i \leq (n-1)/2)}\prod_{k_1^i<k_1^j}\frac{1}{1-x_{k_1^i}x_{k_1^j}}=\prod_{k\leq l}^{(n-1)/2} \prod_{i< j} \frac{1}{1-q^{-1}_{[k,\frac{n-1}{2}]}q^{-1}_{[\frac{n-1}{2},l]}q^{i+j-1}}\\
&=\prod_{k=1}^{(n-1)/2} \prod_{i< j} \frac{1}{1-q_{[k,\frac{n-1}{2}]}^{-2}q^{i+j-1}}  \prod_{k< l}^{(n-1)/2}  \prod_{i< j} \frac{1}{1-q^{-1}_{[k,\frac{n-1}{2}]}q^{-1}_{[\frac{n-1}{2},l]}q^{i+j-1}}\,.
\end{split}
\eq
Combining \req{6factorA} with the second factor of \req{6factorB} gives the second factor of \req{Zrealoddprod}, while the first factor of 
\req{6factorB} combines with the first factor of \req{firstpart} to
\beq
\prod_{j=1}^{(n-1)/2}M^{real}_\sigma(q^{-2}_{[j,\frac{n-1}{2}]},q)\,.
\eq
The ninth factor can be evaluated similarly and yields the same terms as the sixth factor under the replacement $q_k\rightarrow q_k^{-1}$.

The second factor of \req{secondpart} yields
\beq
\prod_{j=1}^{(n-1)/2} \prod_{k_0<k_1^j}\frac{1}{1-x_{k_0} x_{k_1^j}}=\prod_{k=1}^{(n-1)/2}\prod_{i<j}^\infty\frac{1}{1-q^{-1}_{[k,\frac{n-1}{2}]}q^{i+j-1}}\,.
\eq
The seventh factor of \req{secondpart} gives a similar expression with $q_k^{-1}\rightarrow q_k$. The third factor can be evaluated as
\beq\eqlabel{3and4factor}
\prod_{j=1}^{(n-1)/2} \prod_{k_0^1<k_2^j}\frac{1}{1-x_{k_2^j} x_{k_0^1}}=\prod_{k=1}^{(n-1)/2}\prod_{i\leq j} \frac{1}{1-q_{[k,\frac{n-1}{2}]}q^{i+j-1}}\,.
\eq
The fourth factor gives a similar factor with $q_k\rightarrow q_k^{-1}$. We combine these four factors to
\beq\eqlabel{2734factor}
\prod_{k=1}^{(n-1)/2}M(q_{[k,\frac{n-1}{2}]},q)M(q^{-1}_{[k,\frac{n-1}{2}]},q)\,.
\eq
The fifth factor yields
\beq\label{5factor}
\prod_{i,j=1}^{(n-1)/2} \prod_{k_1^i<k_2^j}\frac{1}{1-x_{k_1^i}x_{k_2^j}}=\prod_{k,l=1}^{(n-1)/2} \prod_{i\leq j} \frac{1}{1-q^{-1}_{[k,\frac{n-1}{2}]}q_{[\frac{n-1}{2},\frac{n+1}{2}-l]}q^{i+j-1}}\,.
\eq
Similarly, the eighth factor results in
\beq\label{8factor}
\prod_{i,j=1}^{(n-1)/2} \prod_{k_2^i<k_1^j}\frac{1}{1-x_{k_2^i}x_{k_1^j}}=\prod_{k,l=1}^{(n-1)/2} \prod_{i<j} \frac{1}{1-q^{-1}_{[k,\frac{n-1}{2}]}q_{[\frac{n-1}{2},\frac{n+1}{2}-l]}q^{i+j-1}}\,.
\eq
We combine \req{5factor} and \req{8factor} to obtain
\beq
M(1,q)^{(n-1)/2}\prod_{1\leq i\leq j<(n-1)/2} M(q_{[i,j]},q) M(q^{-1}_{[i,j]},q)\,.
\eq
The first factor can be identified with the constant map contribution and the second factor after combination with \req{2734factor} yields the first factor of \req{Zrealoddprod} plus a similar factor with $q_k\rightarrow q_k^{-1}$.

In summary, we deduce that the real partition functions at the orbifold point for $n$ odd are given by the substitutions \req{closedsub} and \req{realsub} into \req{Zrealoddprod}. Furthermore, we infer that the degree $0$ contribution is given by \req{constmap}.

\paragraph{$n$ even}

It remains to discuss the case with $n$ even. For $n$ even, $q_0$ and $q_{n/2}$ are fixed under identification of colors. We obtain
\beq\eqlabel{Pneven}
\begin{split}
P^{real}_{n\,{\rm even},\sigma}=&\sum_\lambda (-1)^{\frac{(1-\sigma) |\lambda|}{2}}\\
&\times s_\lambda\left(q^{-1}_{[1,\frac{n}{2}-1]}q^{-1/2}_{n/2}q^{1/2},q^{-1}_{[2,\frac{n}{2}-1]}q^{-1/2}_{n/2}q^{1/2},\dots,q^{-1/2}_{n/2}q^{1/2},\right.\\ 
&\left.q^{1/2}_{n/2}q^{1/2}, q_{[\frac{n}{2}-1,\frac{n}{2}-1]} q^{1/2}_{n/2} q^{1/2},q_{[\frac{n}{2}-1,\frac{n}{2}-2]}q^{1/2}_{n/2}q^{1/2},\dots,q_{[\frac{n}{2}-1,1]} q^{1/2}_{n/2}q^{1/2}, \right.\\
&\left. q^{-1}_{[1,\frac{n}{2}-1]}q^{-1/2}_{n/2}q^{3/2},q^{-1}_{[2,\frac{n}{2}-1]},q^{-1/2}_{n/2}q^{3/2},\dots,q^{-1/2}_{n/2}q^{3/2},\right.\\
&\left.q^{1/2}_{n/2}q^{3/2}, q_{[\frac{n}{2}-1,\frac{n}{2}-1]} q^{1/2}_{n/2}q^{3/2},q_{[\frac{n}{2}-1,\frac{n}{2}-2]} q^{3/2},\dots,q_{[\frac{n}{2}-1,1]} q^{3/2},\right.\\
&\left.\phantom{\sum} \dots\right )\,,
\end{split}
\eq
with $q$ now given by $q=q_0^{1/2}q^2_{[1,\frac{n}{2}-1]}q_{n/2}^{1/2}$. 

Similar as in the $n$ odd case discussed above, the first factor of the Schur function identity \req{SchurIDo3} yields
 \beq\eqlabel{epart1}
 \begin{split}
 \prod_{k=1}^{n/2-1}\prod_{i=1}^\infty&\left(\frac{1}{1\pm q_{[k,\frac{n}{2}-1]} q^{1/2}_{n/2} q^{i-1/2}}\right)\left(\frac{1}{1\pm q^{-1}_{[k,\frac{n}{2}-1]} q^{-1/2}_{n/2} q^{i-1/2}}\right)\\
 &\times \left(\frac{1}{1\pm q^{1/2}_{n/2}q^{i-1/2}}\right)  \left(\frac{1}{1\pm q^{-1/2}_{n/2}q^{i-1/2}}\right) \,.
 \end{split}
 \eq

In order to evaluate the second factor of \req{SchurIDo3} we define $K_0=\frac{n}{2}$, $K_0=\frac{n}{2}+1$, $K_2=\{1,\dots,\frac{n}{2}-1\}$ and $K_3=\{\frac{n}{2}+2,\dots,n\}$ and let $K_i^m$ denote the $m$-th element of $K_i$. As before, we define indices $k_i^m$ and ${k'}_i^m$ running over $\{K_i^m,2K_i^m,\dots\}$. Then,
\beq\eqlabel{epart2}
\begin{split}
\prod_{i<j}&=\left(\prod_{k_0^1<{k'}_0^1} \right)\left(\prod_{k_0^1<{k'}_1^1} \right)\left(\prod_{j=1}^{n/2 -1} \prod_{k_0^1<k_2^j}\right)\left(\prod_{j=1}^{n/2 -1} \prod_{k_0^1<k_3^j}\right)   \\
&\times\left(\prod_{k_1^1<{k'}_0^1} \right)\left(\prod_{k_1^1<{k'}_1^1} \right)\left(\prod_{j=1}^{n/2 -1} \prod_{k_1^1<k_2^j}\right)\left(\prod_{j=1}^{n/2 -1} \prod_{k_1^1<k_3^j}\right)   \\
&\times\left( \prod_{j=1}^{n/2-1}\prod_{k^j_2<k_0^1} \right) \left( \prod_{j=1}^{n/2-1}\prod_{k^j_2<k_1^1}  \right)\left( \prod_{i,j=1}^{n/2-1}\prod_{k^i_2<k_2^j}  \right)\left( \prod_{i,j=1}^{n/2-1}\prod_{k^i_2<k_3^j}  \right)\\
&\times\left( \prod_{j=1}^{n/2-1}\prod_{k^j_3<k_0^1} \right) \left( \prod_{j=1}^{n/2-1}\prod_{k^j_3<k_1^1}  \right)\left( \prod_{i,j=1}^{n/2-1}\prod_{k^i_3<k_2^j}  \right)\left( \prod_{i,j=1}^{n/2-1}\prod_{k^i_3<k_3^j}  \right)\,,\\
\end{split}
\eq
and we have to evaluate sixteen factors.

The sixth factor yields
\beq
\prod_{k_1^1<{k'}_1^1}\frac{1}{1-x_{k_1^1}x_{{k'}_1^1}}=\prod_{i<j}\frac{1}{1-q^{-1}_{n/2} q^{i+j-1}}\,.
\eq
We combine it with the third factor of \req{epart1} to a real MacMohan $M^{real}(q^{1/2}_{n/2},q)$ which we identify with the fourth factor in \req{Zrealevenprod}. Similarly, the first factor of \req{epart2} combines with the fourth factor of \req{epart2} to $M^{real}(q^{-1/2}_{n/2},q)$.

The second factor of \req{epart2} gives
\beq
\prod_{k_0^1<{k'}_1^1}\frac{1}{1-x_{k_0^1}x_{{k'}_1^1}}=\prod_{i\leq j}\frac{1}{1-q^{i+j-1}}\,,
\eq
which combines with the fifth factor of \req{epart2} to $M(1,q)$, which we identify with part of the constant map contribution.

The third factor of \req{epart2} results in
\beq
\prod_{j=1}^{n/2 -1} \prod_{k_0^1<k_2^j}=\prod_{k=1}^{n/2 -1}\prod_{i< j}\frac{1}{1-q^{-1}_{[k,\frac{n}{2}]} q^{i+j-1}}\,,
\eq
which we combine with the ninth factor to 
\beq
\prod_{k=1}^{n/2-1} M(q^{-1}_{[k,\frac{n}{2}]},q)\,.
\eq
Similarly, we combine the eights factor with the fourteenth factor to
\beq
\prod_{k=1}^{n/2-1} M(q_{[k,\frac{n}{2}]},q)\,.
\eq
We identify this factor with the first factor in \req{Zrealevenprod}.

The fourth factor yields
\beq
\prod_{j=1}^{n/2 -1} \prod_{k_0^1<k_3^j}=\prod_{k=1}^{n/2 -1}\prod_{i\leq j}\frac{1}{1-q_{[\frac{n}{2}-1,k]}q^{i+j-1}}\,,
\eq
which can be combined with the thirteenth factor to 
\beq\eqlabel{e4and13factor}
\prod_{k=1}^{n/2-1} M(q_{[k,\frac{n}{2}-1]},q)\,.
\eq
Similarly, we combine the seventh with the tenth factor to
\beq
\eqlabel{e7and10factor}
\prod_{k=1}^{n/2-1} M(q^{-1}_{[k,\frac{n}{2}-1]},q)\,.
\eq
The twelfth factor yields
\beq
\prod_{i,j=1}^{n/2 -1} \prod_{k_2^i<k_3^j}=\prod_{k,l=1}^{n/2 -1}\prod_{i\leq j}\frac{1}{1-q_{[k,\frac{n}{2}-1]}q^{-1}_{[\frac{n}{2}-1,l]}q^{i+j-1}}\,.
\eq
The fifteenth factor is similar, but with the second product only over $i<j$. Combination of both factors yields
\beq
M(1,q)^{n/2-1}\prod_{1\leq k\leq l<n/2-1}M(q_{[k,l]},q)M(q_{[k,l]}^{-1},q)\,.
\eq
We identify the first factor with part of the constant map contribution (which is now in total $M(1,q)^{n/2}$) and after combination of the second factor with \req{e4and13factor} and \req{e7and10factor}, it yields the second factor in \req{Zrealevenprod}.

Let us split the eleventh factor into two parts with $i< j$ and $i\geq j$, \ie,
\beq\eqlabel{e11factora}
\prod_{1\leq i<j\leq n/2-1} \prod_{k^i_2<k^j_2}\frac{1}{1-x_{k^i_2}x_{k^j_2}}=\prod_{k<l}^{n/2-1}\prod_{i\leq j} \frac{1}{1-q^{-1}_{[k,n/2]} q^{-1}_{[l,n/2-1]} q^{i+j-1}}\,,
\eq
and
\beq\eqlabel{e11factorb}
\prod_{1 \leq j\leq i}^{n/2-1} \prod_{k^i_2<k^j_2}\frac{1}{1-x_{k^i_2}x_{k^j_2}}=\prod_{k=1}^{n/2-1}\prod_{i<j}\frac{1}{1-q^{-2}_{[k,\frac{n-2}{2}]}q^{-1}_{n/2}q^{i+j-1}}\prod_{k<l}^{n/2-1}\prod_{i<j}\frac{1}{1- q^{-1}_{[k,\frac{n}{2}]}q^{-1}_{[\frac{n-2}{2},l]} q^{i+j-1}}\,.
\eq
We combine the first factor of \req{e11factorb} with the second factor of \req{epart1} to 
\beq
\prod_{k=1}^{n/2-1}M^{real}_\pm(q^{-2}_{[k,\frac{n-2}{2}]}q^{-1}_{n/2},q)\,,
\eq
and the second factor of \req{e11factorb} with \req{e11factora} to 
\beq
\prod_{1\leq i<j<n/2}M(q^{-1}_{[i,n/2]}q^{-1}_{[j,n/2-1]},q)\,.
\eq
The sixteenth factor can be evaluated in a similar fashion and we obtain from it the third and fifth factor of \req{Zrealevenprod}.

Collecting all factors, we deduce that the real partition functions at the orbifold point for $n$ even are given by the substitutions \req{closedsub} and \req{realsub} into \req{Zrealevenprod}. Furthermore, we infer that the degree $0$ contribution is given by \req{constmap}.

\subsection{Derivation of real conifold length $N$ pyramid partition function}
\label{LNconicalc}
In this subsection we evaluate the correlators \req{RconiNodd0} and \req{RconiNeven0}. Let us start with \req{RconiNodd0}.

\paragraph{$N-1$ even}
We can either let the $\Gamma'_+$ operators act on the left state and use the relation \req{statetoSchur}, or we can commute them to the right via the relations \req{Gcommute} such that they are annihilated by the vacuum state. The latter is more convenient. We infer
\beq\eqlabel{RconiNodd}
\begin{split}
R_{N\,{\rm odd},\sigma}=&\prod_{i\leq j}^{(N-1)/2}\frac{1}{1+v_iu_j^{-1}}\prod_{i,j}^{(N-1)/2,\infty}(1+u_i^{-1}v_j)\prod_{i,j}^{(N-1)/2,\infty}\frac{1}{1-xu_i^{-1}u_j}   \\
&\times R_\sigma(u_i\rightarrow x\cdot u_i,v_i)\,,
\end{split}
\eq
with $R_\sigma(u_i\rightarrow x\cdot u_i,v_i)$ as in \req{Rconiprefinal}. We rewritte the second factor of \req{RconiNodd} as
\beq\eqlabel{RconiNoddfac2}
\prod_{i,j}^{(N-1)/2,\infty}(1+u_i^{-1}v_j)=\prod_{i=1}^\infty(1+q_1^{-1}q^i)^{(N-1)/2}\prod_{i\geq j}^{(N-1)/2}(1+u_i^{-1}v_j)\,.
\eq
Observe that the second factor of \req{RconiNoddfac2} cancels against the first factor of \req{RconiNodd}, while the first factor is a part of $M^{real}_{N,\pm}(-q_1^{-1} q^{-(N-1)},q)^{-1}$ (\cf, \req{MNrealfull}). From \req{Rconiprefinal} we infer that the $R_\sigma(u_i\rightarrow x\cdot u_i,v_i)$ factor yields $M_\pm^{real}(-q_1 x^2,q)^{-1}$, $M_\pm^{real}(-q_1^{-1},q)^{-1}$ and the second factor of  \req{Rconiprefinal} combines with the third factor of \req{RconiNodd} to the degree $0$ contribution $M(1,q)$. In summary, we obtain \req{Rconifinal}. 

\paragraph{$N-1$ odd}
As in the $N-1$ even case, we commute the $\Gamma_+'$ operators to the right and obtain
\beq\eqlabel{RconiNeven}
\begin{split}
R_{N\,{\rm even},\sigma}=&\prod_{i\leq j}^{N/2-1}\frac{1}{1+v_iu_j^{-1}}\prod_{i,j}^{N/2,\infty}(1+qu_i^{-1}v_j)\prod_{i,j}^{N/2,\infty}\frac{1}{1+xq^{1/2}u_i^{-1}u_j}\\
&\times R_\sigma(u_i\rightarrow x\cdot u_i,v_i\rightarrow q^{1/2} v_i)\,.
\end{split}
\eq
We rewrite the second factor of \req{RconiNeven} as
\beq\eqlabel{RconiNevenfac2}
\prod_{i,j}^{N/2,\infty}(1+qu_i^{-1}v_j)=\prod_{i=1}^\infty(1+q_1^{-1}q q^i)^{N/2}\prod^{N/2-1}_{i\geq j}(1+u_i^{-1}v_j)\,.
\eq
Note that the second factor of \req{RconiNevenfac2} cancels against the first factor of \req{RconiNeven} while the first factor is a part of 
$M_{N,\pm}^{real}(-q_1^{-1}q^{-(N-1)},q)^{-1}$. Similar as in the $N-1$ even case, the $R_\sigma(u_i\rightarrow x\cdot u_i,v_i\rightarrow q^{1/2} v_i)$ factor yields $M^{real}_\pm (-q_1^{-1}q,q)^{-1}$, $M^{real}_\pm(-x^2 q_1,q)^{-1}$ and a factor which combines with the third factor of \req{RconiNeven} to the degree 0 contribution. Hence, we obtain \req{Rconifinal} as well for $N-1$ odd.

\subsection{Derivation of real $\C^2/\Z_2\times\C$ length $N$ pyramid partition function}
\label{C2Z2correv}
In this section we evaluate \req{C2Z2lNoddcorr} and \req{C2Z2lNevencorr}. This is very similar to the correlators evaluated in section \ref{LNconicalc}. Therefore we will be brief.

\paragraph{$N-1$ even}
Commuting the $\Gamma_+$ operators to the right yields
\beq
\begin{split}
R_{N\,{\rm odd},\sigma}&=\prod_{i\leq j}^{(N-1)/2}(1-v_iu_j^{-1})\prod_{i,j}^{(N-1)/2,\infty}\frac{1}{1-  u_i^{-1}v_j}\prod_{i,j}^{(N-1)/2,\infty}\frac{1}{1-x u_i^{-1}u_j}\\
&\times R_\sigma(u\rightarrow x\cdot u,v)\,.
\end{split}
\eq
with $R_\sigma(u\rightarrow x\cdot u,v)$ as in \req{RC2Z2final}. Comparing with \req{RconiNodd}, we immediately deduce that we can obtain in a similar fashion \req{C2Z2lengthNfinal}.

\paragraph{$N-1$ odd}
Similar as above, we obtain
\beq
\begin{split}
R_{N\,{\rm even},\sigma}&=\prod_{i\leq j}^{N/2-1}(1-v_iu_j^{-1})\prod_{i,j}^{N/2,\infty}\frac{1}{1- q u_i^{-1}v_j}\prod_{i,j}^{N/2,\infty}\frac{1}{1-x q^{1/2} u_i^{-1}u_j}\\
&\times R_\sigma(u\rightarrow x\cdot u,v\rightarrow q^{1/2} v)\,,
\end{split}
\eq
from which we again deduce that \req{C2Z2lengthNfinal} holds.


\begin{thebibliography}{99}
\addcontentsline{toc}{section}{References}
\renewcommand{\itemsep}{-.2cm}
\colorlinksblue
\small

\bibitem{Walcher:2007qp}
  J.~Walcher,
  ``Evidence for Tadpole Cancellation in the Topological String,''
  \arxiv{0712.2775}{hep-th}.

\bibitem{Krefl:2009md}
  D.~Krefl and J.~Walcher,
  ``The Real Topological String on a local Calabi-Yau,''
  \arxiv{0902.0616}{hep-th}.

\bibitem{Krefl:2009mw}
  D.~Krefl, S.~Pasquetti and J.~Walcher,
  ``The Real Topological Vertex at Work,''
  \arxiv{0909.1324}{hep-th}.


\bibitem{Sinha:2000ap}
  S.~Sinha and C.~Vafa,
  ``SO and Sp Chern-Simons at large N,''
  \hepth{0012136}.

\bibitem{Acharya:2002ag}
  B.~S.~Acharya, M.~Aganagic, K.~Hori and C.~Vafa,
  ``Orientifolds, mirror symmetry and superpotentials,''
  \hepth{0202208}.

\bibitem{Bouchard:2004iu}
  V.~Bouchard, B.~Florea and M.~Marino,
  ``Counting higher genus curves with crosscaps in Calabi-Yau orientifolds,''
  JHEP {\bf 0412} (2004) 035
  \hepth{0405083}.

\bibitem{Bouchard:2004ri}
  V.~Bouchard, B.~Florea and M.~Marino,
  ``Topological open string amplitudes on orientifolds,''
  JHEP {\bf 0502} (2005) 002
  \hepth{0411227}.


\bibitem{Cook:2008eu}
  P.~L.~H.~Cook, H.~Ooguri and J.~Yang,
  ``New Anomalies in Topological String Theory,''
  Prog.\ Theor.\ Phys.\ Suppl.\  {\bf 177} (2009) 120
  \arxiv{0804.1120}{hep-th}.

\bibitem{Bonelli:2009aw}
  G.~Bonelli, A.~Prudenziati, A.~Tanzini and J.~Yang,
  ``Decoupling A and B model in open string theory -- Topological adventures in
  the world of tadpoles,''
  JHEP {\bf 0906} (2009) 046
  \arxiv{0905.1286}{hep-th}.

\bibitem{Aganagic:2000gs}
  M.~Aganagic and C.~Vafa,
  ``Mirror symmetry, D-branes and counting holomorphic discs,''
  \hepth{0012041}.


\bibitem{Aganagic:2003db}
  M.~Aganagic, A.~Klemm, M.~Marino and C.~Vafa,
  ``The topological vertex,''
  Commun.\ Math.\ Phys.\  {\bf 254} (2005) 425
  \hepth{0305132}.

\bibitem{Okounkov:2003sp}
  A.~Okounkov, N.~Reshetikhin and C.~Vafa,
  ``Quantum Calabi-Yau and classical crystals,''
  \hepth{0309208}.


\bibitem{Iqbal:2003ds}
  A.~Iqbal, N.~Nekrasov, A.~Okounkov and C.~Vafa,
  ``Quantum foam and topological strings,''
  JHEP {\bf 0804} (2008) 011
  \hepth{0312022}.

\bibitem{MNOPI}
D.~Maulik, N.~Nekrasov, A.~Okounkov and R.~Pandharipande,
``Gromov-Witten theory and Donaldson-Thomas theory, I"
\math{0312059}.

\bibitem{Denef:2007vg}
  F.~Denef and G.~W.~Moore,
  ``Split states, entropy enigmas, holes and halos,''
  \hepth{0702146}.

\bibitem{Young08}
B.~Young and J.~Bryan,
``Generating functions for colored 3D Young diagrams and the Donaldson-Thomas invariants of orbifolds,"
\arxiv{0802.3948}{math.CO}.

\bibitem{Szendroi07}
B.~Szendroi, ``Non-commutative Donaldson-Thomas theory and the conifold,"
\arxiv{0709.3419}{math.AG}

\bibitem{Mozgovoy:2008fd}
  S.~Mozgovoy and M.~Reineke,
  ``On the noncommutative Donaldson-Thomas invariants arising from brane
  tilings,''
  \arxiv{0809.0117}{math.AG}.

\bibitem{Aganagic:2009kf}
  M.~Aganagic, H.~Ooguri, C.~Vafa and M.~Yamazaki,
  ``Wall Crossing and M-theory,''
  \arxiv{0908.1194}{hep-th}.

\bibitem{MNOPII}
D.~Maulik, N.~Nekrasov, A.~Okounkov and R.~Pandharipande,
``Gromov-Witten theory and Donaldson-Thomas theory, II"
\math{0406092}.

\bibitem{Walcher:2006rs}
  J.~Walcher,
  ``Opening mirror symmetry on the quintic,''
  Commun.\ Math.\ Phys.\  {\bf 276} (2007) 671
  \hepth{0605162}.

\bibitem{Hori:2005bk}
  K.~Hori, K.~Hosomichi, D.~C.~Page, R.~Rabadan and J.~Walcher,
  ``Non-perturbative orientifold transitions at the conifold,''
  JHEP {\bf 0510} (2005) 026
  \hepth{0506234}.

\bibitem{Karp:2005vq}
  D.~Karp, C.~C.~Liu and M.~Marino,
  ``The local Gromov-Witten invariants of configurations of rational curves,''
  \math{0506488}.

\bibitem{Bryan08}
J.~Bryan and A.~Gholampour
``The Quantum McKay Correspondence for polyhedral singularities,"
\arxiv{0803.3766}{math.AG}.

\bibitem{Katz:1999xq}
  S.~H.~Katz, A.~Klemm and C.~Vafa,
  ``M-theory, topological strings and spinning black holes,''
  Adv.\ Theor.\ Math.\ Phys.\  {\bf 3} (1999) 1445
  \hepth{9910181}.

\bibitem{Klemm:2004km}
  A.~Klemm, M.~Kreuzer, E.~Riegler and E.~Scheidegger,
  ``Topological string amplitudes, complete intersection Calabi-Yau spaces  and
  threshold corrections,''
  JHEP {\bf 0505} (2005) 023
  \hepth{0410018}.

\bibitem{Gopakumar:1998ii}
  R.~Gopakumar and C.~Vafa,
  ``M-theory and topological strings. I,''
  \hepth{9809187}.

\bibitem{Gopakumar:1998jq}
  R.~Gopakumar and C.~Vafa,
  ``M-theory and topological strings. II,''
  \hepth{9812127}.

\bibitem{Behrend05}
K.~Behrend,
``Donaldson-Thomas type invariants via microlocal geometry,"
\mathag{0507523}

\bibitem{BF05}
K.~Behrend and B.~Fantechi,
``Symmetric Obstruction Theories and Hilbert Schemes of Points on Threefolds,"
\math{0512556}.

\bibitem{OR05}
A.~Okounkov, N. Reshetikhin,
``Random skew plane partitions and the Pearcey process,"
\math{0503508}.

\bibitem{Kac}
V.~G.~Kac,
``Infinite dimensional Lie algebras,"
third edition, Cambridge University Press, 1990

\bibitem{Sulkowski:2009rw}
  P.~Sulkowski,
  ``Wall-crossing, free fermions and crystal melting,''
  \arxiv{0910.5485}{hep-th}.

\bibitem{Franco:2007ii}
  S.~Franco, A.~Hanany, D.~Krefl, J.~Park, A.~M.~Uranga and D.~Vegh,
  ``Dimers and Orientifolds,''
  JHEP {\bf 0709} (2007) 075
  \arxiv{0707.0298}{hep-th}.

\bibitem{Kennaway:2007tq}
  K.~D.~Kennaway,
  ``Brane Tilings,''
  Int.\ J.\ Mod.\ Phys.\  A {\bf 22} (2007) 2977
  \arxiv{0706.1660}{hep-th}.

\bibitem{Bergh}
M.~Van~den~Bergh, ``Non-commutative crepant resolutions," The legacy of Niels Henrik Abel, 749-770, Springer, 
Berlin, 2004. 

\bibitem{King}
A.~King, ``Moduli representations of finite-dimensional algebras," 
Quart. J. Math. Oxford Ser. (2) 45(180):515-530, 1994.

\bibitem{Ooguri:2008yb}
  H.~Ooguri and M.~Yamazaki,
  ``Crystal Melting and Toric Calabi-Yau Manifolds,''
  Commun.\ Math.\ Phys.\  {\bf 292} (2009) 179
\arxiv{0811.2801}{hep-th}.

\bibitem{Young07}
B.~Young,
``Computing a pyramid partition generating function with dimer shuffling,"
\arxiv{0709.3079}{math.CO}.

\bibitem{Chuang:2008aw}
  W.~y.~Chuang and D.~L.~Jafferis,
  ``Wall Crossing of BPS States on the Conifold from Seiberg Duality and
  Pyramid Partitions,''
  Commun.\ Math.\ Phys.\  {\bf 292} (2009) 285
  \arxiv{0810.5072}{hep-th}.

\bibitem{Nagao08}
K.~Nagao and H.~Nakajima,
``Counting invariant of perverse coherent sheaves and its wall-crossing,"
\arxiv{0809.2992}{math.AG}.

\bibitem{Aganagic:2009cg}
  M.~Aganagic and M.~Yamazaki,
  ``Open BPS Wall Crossing and M-theory,''
  \arxiv{0911.5342}{hep-th}.

\bibitem{Jafferis:2008uf}
  D.~L.~Jafferis and G.~W.~Moore,
  ``Wall crossing in local Calabi Yau manifolds,''
  \arxiv{0810.4909}{hep-th}.

\bibitem{Ooguri:1999bv}
  H.~Ooguri and C.~Vafa,
  ``Knot invariants and topological strings,''
  Nucl.\ Phys.\  B {\bf 577} (2000) 419
  \hepth{9912123}.

\bibitem{Denef:2009ja}
  F.~Denef, M.~Esole and M.~Padi,
  ``Orientiholes,''
  \arxiv{0901.2540}{hep-th}.

\bibitem{Ooguri:2004zv}
  H.~Ooguri, A.~Strominger and C.~Vafa,
  ``Black hole attractors and the topological string,''
  Phys.\ Rev.\  D {\bf 70} (2004) 106007
  \hepth{0405146}.

\bibitem{Aganagic:2004js}
  M.~Aganagic, H.~Ooguri, N.~Saulina and C.~Vafa,
  ``Black holes, q-deformed 2d Yang-Mills, and non-perturbative topological
  strings,''
  Nucl.\ Phys.\  B {\bf 715} (2005) 304
  \hepth{0411280}.

\bibitem{Aganagic:2005dh}
  M.~Aganagic, A.~Neitzke and C.~Vafa,
  ``BPS microstates and the open topological string wave function,''
  \hepth{0504054}.

\bibitem{Stanley}
R.~P.~Stanley,
``Enumerative Combinatorics,"
Vol. 2, Cambridge University Press, 1992

\bibitem{MacDonald}
I.~G.~MacDonald,
``Symmetric Functions and Hall Polynomials,"
second edition, Oxford University Press, 1999





















\end{thebibliography}
\end{document}